\documentclass[aip,fleqn,layout=traditional,journal=jpccck,article]{achemso}
\usepackage[version=3]{mhchem} % Formula subscripts using \ce{}
\usepackage{mciteplus}
\usepackage{multirow}
\usepackage{arydshln}
\usepackage{amsfonts,calrsfs}
\usepackage{braket}
\usepackage{amsmath}

\newcommand{\Tr}{\operatorname{Tr}}

\newcommand{\onlinecite}[1]{\!\citenum{#1}}

\author{Peter Schwerdtfeger}
\email{peter.schwerdtfeger@gmail.com}
\affiliation[NZIAS Massey University]
{Centre for Theoretical Chemistry and Physics, The New Zealand Institute for Advanced Study (NZIAS), Massey University Albany, Private Bag 102904, Auckland 0745, New Zealand}
\author{Antony Burrows}%\email{a.burrows1@massey.ac.nz}
\affiliation[NZIAS Massey University]
{Centre for Theoretical Chemistry and Physics, The New Zealand Institute for Advanced Study (NZIAS), Massey University Albany, Private Bag 102904, Auckland 0745, New Zealand}
\author{Odile R. Smits}%\email{o.smits@massey.ac.nz}
\affiliation[NZIAS Massey University]
{Centre for Theoretical Chemistry and Physics, The New Zealand Institute for Advanced Study (NZIAS), Massey University Albany, Private Bag 102904, Auckland 0745, New Zealand}

\title[ELJ] {The Lennard Jones Potential Revisited -- Analytical Expressions for Vibrational Effects in Cubic and Hexagonal Close-Packed Lattices}
\begin{document}


%{\bf Odile: Does it make more sense to use for the solid state equilibrium distance $r_0$ or $r_s^{min}$ (instead of $r_0^{min}$), and for the nearest neighbor distance $r_s$ instead of $r_0$. My brain keeps on being confused: I keep on thinking $r_0$ stands for solid state equilibrium distance... What do you use in other published articles? I see in the original ELJ paper $r^{SS}$ was used for solid state nearest neighbor distance, and $r^{ss}_{min}$ for ss equilibrium distance. But that might also be confusing to use in this article.. } 

\clearpage
\date{\today}
\begin{abstract}
Analytical formulae are derived for the zero-point vibrational energy and anharmonicity
corrections of the cohesive energy and the mode Gr\"{u}neisen parameter within the Einstein model for the 
cubic lattices (sc, bcc and fcc) and for the hexagonal close-packed structure.
This extends the work done by Lennard Jones and Ingham in 1924, Corner in 1939 and Wallace in 1965.
The formulae are based on the description of two-body energy contributions by an 
inverse power expansion (extended Lennard-Jones potential). These make use of three-dimensional lattice sums, 
which can be transformed to fast converging series and accurately determined by various expansion 
techniques. 
We apply these new lattice sum expressions to the rare gas solids and discuss associated critical points. 
The derived formulae give qualitative but nevertheless deep insight into vibrational effects in solids from the lightest (helium) 
to the heaviest rare gas element (oganesson), both presenting special cases because of strong quantum effects for the 
former and strong relativistic effects for the latter. 
\end{abstract}


\begin{figure}[htbp]
\begin{center}
\includegraphics[width=9cm]{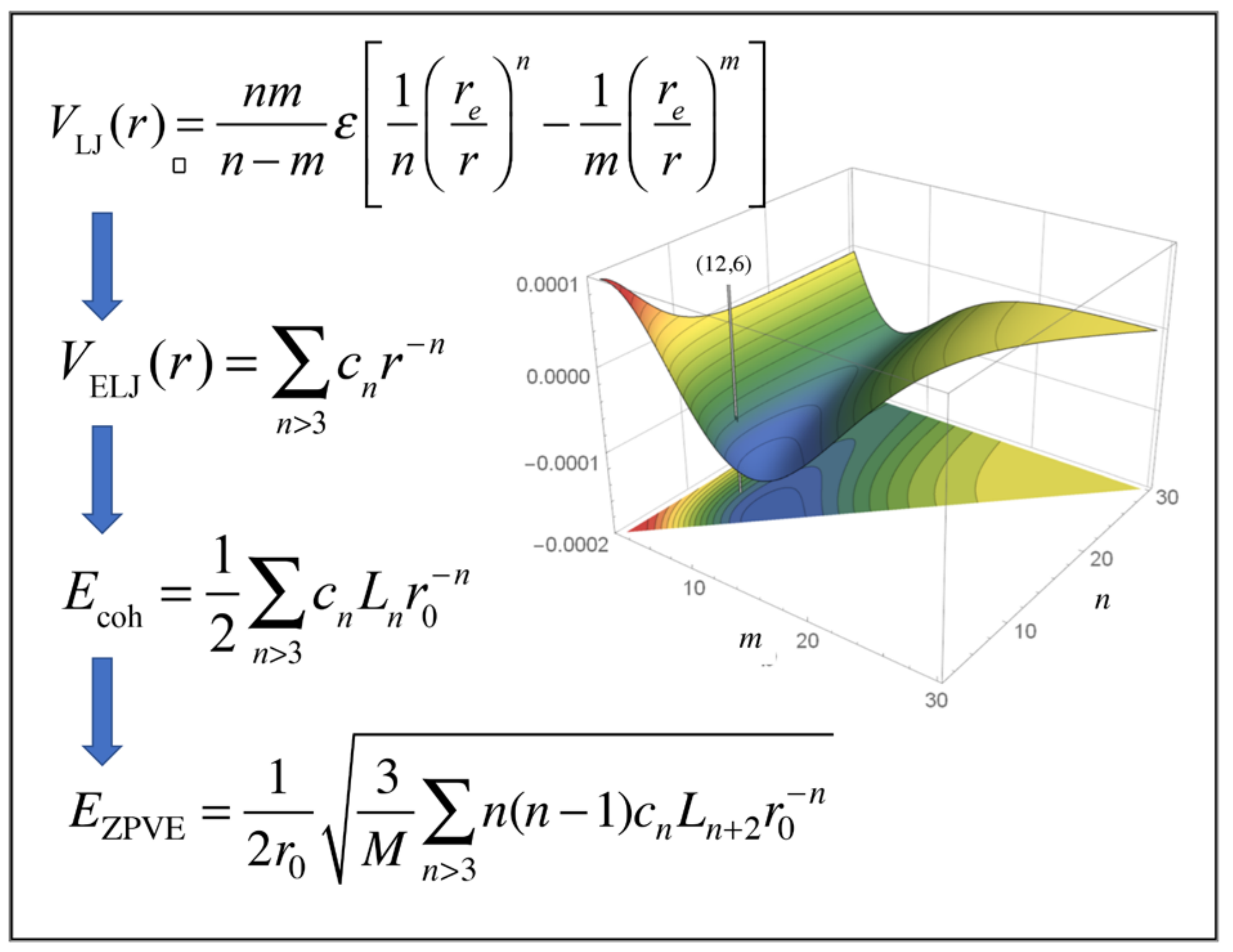}
\end{center}
\end{figure}

\clearpage

%Introduction
\section{\label{sec:intro}Introduction}

The $(n,m)$ Lennard-Jones (LJ) potential\cite{Jones-1924-0,Jones1924a,Simon-1924,Jones-1931,Lennard-Jones1937} is, beside the Morse potential \cite{Morse1929}, the most widely used interaction potential in the physical and biological sciences,\cite{Kollman2004,Stephan2019,Stephan2020}
\begin{equation} \label{eq:LJ}
V_{\rm LJ}(r)=\frac{nm}{n-m}\epsilon \left[ \frac{1}{n}\left( \frac{r_e}{r} \right)^{n} - \frac{1}{m} \left( \frac{r_e}{r} \right)^{m} \right]
\end{equation}
with an equilibrium distance $r_e$ and binding energy $\epsilon$ (taken as a positive value) between two interacting systems.

The story of how this interaction potential came to be commonly known today as the LJ potential started with Mie's 1903 discussion suggesting an equation of state containing a volume dependent term of the form $(AV^{-1}-BV^{-\nu/3})$ with $\nu>3$.\cite{Mie-1903} Following this, in 1912 Gr\"uneisen\cite{Gruneisen1912} published the exact formula for what became the well known $(n,m)$ LJ potential, and in 1920 Krater also introduced a less general (2,1) potential which went unnoticed.\cite{Kratzer-1920} The Gr\"uneisen $(n,m)$ potential was modified by Born and Land\'e\cite{born1918} in 1918 for ionic crystal and the same year Mandelung introduced the lattice sum for ionic crystals today known as the Mandelung constant.\cite{madelung1918} It wasn’t until 1924 after Lennard-Jones solved the equation of state analytically to derive the parameters based on experimental results, that the LJ $(n,m)$ potential gained notoriety.\cite{Jones-1924-0} However the physical relevance of the long-rang dispersive term came much later in 1930 by London.\cite{London-1930} What is curious about the chronology is that Simon and Simpson used the Gr\"uneisen potential in 1924 giving it a proper citation, and Lennard-Jones in his second paper also cited Simon and Simpsons paper in 1924 within a series of papers, but Gr\"uneisen's paper was ignored.

To allow for a more accurate description of the interacting potential, the LJ potential has been generalized into an inverse power series of the form\cite{born1940,Schwerdtfeger-2006}
\begin{equation} \label{eq:ELJ}
V_{\rm ELJ}(r)=\sum_{n=1}^{n_\textrm{max}} c_nr^{-s_n},
\end{equation}
with $c_n\in\mathbb{R}$ and $s_n\in\mathbb{R_+}$ ($s_1$=6 and $s_2$=12 for the (12,6) LJ potential). A boundary condition such that the minimum is positioned at a distance $r_e$ with a potential depth $\epsilon$  $\sum_{n=1}^{n_\textrm{max}} c_nr_e^{-s_n}=-\epsilon$ with $\epsilon >0$. The coefficients $c_n$ can be obtained from either experimental data or accurate quantum-theoretical calculations.\cite{Schwerdtfeger-2006,Pahl_JChemPhys132}. The advantage of the inverse power series compared to more complicated expressions like the Morse potential,\cite{Morse1929} or accurate potential forms separating the long-range from the short-range region,\cite{Tang-Toennis-1984,Jager_MolPhys107,PatkowskiSzalewicz2010} is that one can express analytically the volume dependent two-body (static) cohesive energy of certain lattices in terms of infinite lattice sums,
\begin{equation} \label{eq:VolLJ}
\begin{aligned}
E_{\rm ELJ}(V) = \lim_{N\rightarrow \infty}\frac{1}{N} \sum_{i<j}^N V_{\rm ELJ}(r_{ij})=\frac{1}{2}\sum_{i=1}^\infty V_{\rm ELJ}(r_{0i})\\
=\frac{1}{2}\sum_{n=1}^{n_\textrm{max}} c_nL_{s_n}r_0^{-s_n}=\frac{1}{2}\sum_{n=1}^{n_\textrm{max}} f_L^{s_n/3}c_nL_{s_n}V^{-s_n/3} .
\end{aligned}
\end{equation}
Here, $n>m$ guarantees the existence of a minimum and $s_n>3$ to guarantee convergence for the 3D bulk system.\cite{Schwerdtfeger-2006} In Eq.(\ref{eq:VolLJ}) $r_0$ is the nearest neighbor distance of the lattice $r_0={\rm min}\{r_{0i}\}$, with $r_{0i}$ being the distance from one selected atom in the lattice to all other atoms $i$), and $f_L$ is a lattice-specific parameter converting $r_0$ into the volume $V=f_Lr_0^3$, i.e. $f_{\rm sc}=1$, $f_{\rm bcc}=4/(3\sqrt{3})$, $f_{\rm fcc}= f_{\rm hcp}=1/\sqrt{2}$. We use the fact that for a cubic lattice the summation over all atoms $i$ and $j$ with distance $r_{ij}$ simplifies to summing over all interactions from one selected atom placed at the origin to all other atoms $i$ in the solid because of translational symmetry. Once basic lattice vectors are introduced to express the distances $r_{0i}$ from the chosen atom to all other atoms in the lattice, the cohesive energy can be expressed in terms of three-dimensional lattice sums $L_{s}\in\mathbb{R_+}$ multiplied by inverse powers of the nearest neighbor distance $r_0$ as originally described by Lennard-Jones in 1924\cite{Jones-1924b,Jones-1925} and analyzed in detail by Borwein et al.\cite{borwein-2013} 

For example, the $(n,m)$ LJ potential, and more specifically, the (12,6) LJ potential with coefficients $c_{1}=\epsilon r_e^{12}$ ($s_1=12$) and $c_2=-2\epsilon r_e^6$ ($s_2=6$), becomes (in atomic units),
\begin{equation} \label{eq:EcLJn-m}
\begin{aligned}
E_{\rm LJ}(r_{0}) &=\frac{nm\epsilon}{2(n-m)}\left[\frac{1}{n} L_{n}\left(\frac{r_e}{r_{0}}\right)^{n} - \frac{1}{m} L_m\left(\frac{r_e}{r_{0}}\right)^{m}\right]  \\
&\mathop{=}\limits_{m=6}^{n=12} \quad \epsilon \left(\frac{r_e}{r_{0}}\right)^{6}
       \left[ \frac{L_{12}}{2} \left(\frac{r_e}{r_{0}}\right)^{6} - L_{6}   \right] \,.
\end{aligned}
\end{equation}

From Eq.(\ref{eq:VolLJ}) one easily obtains the corresponding analytical expressions for the volume dependent pressure $P$ and the bulk modulus $B$ of a lattice expressed in terms of lattice sums as,\cite{Schwerdtfeger-2006}
\begin{equation} \label{eq:PLJ}
P_{\rm ELJ}(V)=-\frac{\partial E_{\rm ELJ}(V)}{\partial V}=\frac{1}{6V}\sum_{n=1}^{n_\textrm{max}} s_n c_nL_{s_n}r_0^{-s_n}=\frac{1}{6}\sum_{n=1}^{n_\textrm{max}} s_n f_L^{s_n/3}c_nL_{s_n}V^{-\tfrac{s_n}{3}-1},
\end{equation}
\begin{equation} \label{eq:BLJ}
\begin{aligned}
B_{\rm ELJ}(V)&=V\frac{\partial^2 E_{\rm ELJ}(V)}{\partial V^2}=\frac{1}{18V}\sum_{n=1}^{n_\textrm{max}} s_n\left( s_n+3\right) c_nL_{s_n}r_0^{-s_n} \\ &=\frac{1}{18}\sum_{n=1}^{n_\textrm{max}} s_n\left( s_n+3\right) f_L^{s_n/3}c_nL_{s_n}V^{-\tfrac{s_n}{3}-1} .
\end{aligned}
\end{equation}
These formulae clearly demonstrate the usefulness of an extended LJ potential as important solid-state properties can be calculated {\it analytically} to computer precision for any volume $V$ or pressure $P$ if the lattice sums are accurately known. 

Working on the melting of argon, Herzfeld and Goeppert-Mayer pointed out as early as in 1934 that lattice vibrations increase the equilibrium lattice distance and must therefore be considered.\cite{Herzfeld1934} Corner\cite{Corner-1939} and Wallace\cite{Wallace1965} analyzed such lattice vibrational effects in more detail for the $(n,6)$ LJ potential, and through approximations derived an analytical formula for the zero-point vibrational energy of the fcc lattice. Later, Nijboer and deWette analyzed lattice vibrations in $k$-space for the dynamic matrix for a face-centered cubic crystal with a varying lattice constant.\cite{Nijboer1965,Nijboer1971} However, the corresponding lattice sums become rather complicated, and fast converging forms for the dynamic matrix for phonon dispersion are not available. 

In this paper we derive exact analytical expressions for the zero-point vibrational energy and corresponding anharmonicity correction to the cohesive energy and the lattice (mode) Gr\"{u}neisen parameter within the Einstein approximation.\cite{Einstein1907} That is, moving a single atom in the field of an ELJ potential, for the simple cubic (sc), body-centered cubic (bcc) or face centered cubic (fcc) lattices, including thermodynamic properties, and applying these formulae to various model systems for the rare gases from helium to the heaviest element in this group, oganesson. We also include in our discussion the more complicated hexagonal close-packed structure (hcp). As specific applications we focus on the high-pressure range of helium, and the fcc and hcp phase for argon which are energetically very close, and discuss the limitations of the Einstein model. For the Gr\"{u}neisen parameter we investigate solid neon as an example where anharmonicity effects are large.

%Theoretical Part
\section{\label{sec:Theory}Theoretical Foundations}

%\subsubsection{\label{sec:cubic}The cubic lattices}
The total cohesive energy per atom, $E_{{\mathop{\rm coh}}}(V) $, can be divided into static $E^{{\rm stat}}_{{\rm coh}}(V) $ and dynamic $E^{{\rm dyn}}_{{\rm coh}}(V) $ contributions, the latter resulting from zero-point vibrational motion:
\begin{align} \label{eq:nbody2contributions}
E_{{\mathop{\rm coh}}}(V) =  E^{{\rm stat}}_{{\rm coh}}(V) + E^{{\rm dyn}}_{{\rm coh}}(V) \,.
\end{align}
The total static contribution can be approximated within the many-body ansatz including two- and higher body contributions in the solid if the many-body expansion is converging fast.\cite{N-body.AH2007} We use translational symmetry to evaluate the most important two-body contribution through an ELJ potential,  $E^{{\rm stat}}_{{\rm coh}}(V)\cong E_{\rm ELJ}(V)$, and for the dynamic part, 
\begin{align} \label{eq:nbody2contributionsZPVE}
E^{{\rm dyn}}_{{\rm coh}}(V)\cong E_{\rm ELJ}^{\rm ZPVE}(V)+E_{\rm ELJ}^{\rm AZPVE}(V) \,.
\end{align}
We apply the Einstein approximation for a vibrating atom in the interacting ELJ field of all other atoms.
Here $E_{\rm ELJ}^{\rm ZPVE}(V)$ is the volume dependent zero-point vibrational energy (ZPVE) contribution within the harmonic oscillator approximation, and $E_{\rm ELJ}^{\rm AZPVE}(V)$ is the corresponding anharmonicity correction (AZPVE). Although this treatment neglects important higher-body contributions and phonon dispersion, and for helium important quantum effects originating from the nuclear motion, analytical formulae derived in terms of Eq.(\ref{eq:nbody2contributions}) will provide us with  some useful qualitative insight into solid-state properties. For a more accurate treatment which goes beyond this approximation see Ref.\onlinecite{Schwerdtfeger-2016} for example, where J/mol accuracy has been achieved for the cohesive energy of solid argon.

%\subsection{\label{sec:Lsums}Lattice sums}
\subsection{\label{sec:Lsums}Lattice sums}
Lattice sums are of key importance in the work presented in this article, a field pioneered early on by Lennard-Jones.\cite{Jones-1924b,Jones-1925} 
Any expression in inverse powers of distances for interacting atoms in a lattice can be uniquely described by a three-dimensional lattice sum $L_s$ (if convergent). For the case of the cubic lattices sc, bcc and fcc we have,\cite{borwein-2013}
\begin{equation} \label{eq:mapping}
\sum_{i=1}^\infty r_{i}^{-s}=L_{s}r_0^{-s},
\end{equation}

where the sum runs over all lattice points $i$ in three dimensions located at distances $r_i$ from a selected atom is reduced to $L_s$ multiplied by the nearest neighbor distance, $r_0$ to the power of $s$ ( $s>3$ to ensure convergence of the lattice sum).  

\begin{figure}[t]
\centering
\includegraphics[width=.5\columnwidth]{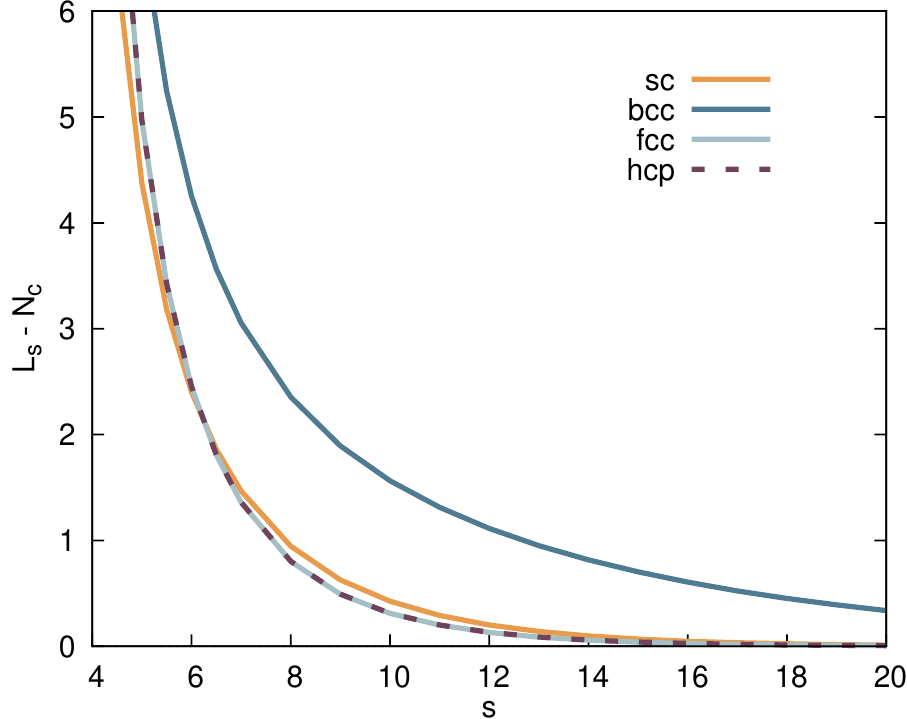}
\caption{ Lattice sums, $L_s$, minus the kissing number, $L_\infty=N_c$, of sc, bcc, fcc and hcp for a range of real exponent $s$.}
\label{fig:LatticeSums}
\end{figure}

Analytical expressions for lattice sums $L_s$ (also called Lennard-Jones-Ingham parameters) have a long history\cite{born-1998,borwein-2013} and have been tabulated for a number of lattices with integer exponents ($s\in \mathbb{N}$) by several authors.\cite{Wallace1965,glasser-zucker-1980,borwein1985,Schwerdtfeger-2006,Stein2017,burrows-2020} Even for more complicated lattices such as hcp, expressions of the cohesive energy in terms of lattice sums have been formulated\cite{burrows-2020} based on the 1940 paper by Kane and Goeppert-Mayer.\cite{Kane-1940} For the lattices considered in this work we have the following lattice sums
\begin{equation} \label{eq:latticesum1}
\begin{aligned}
L_s^{\rm sc} &= {\sum\limits_{i,j,k \in \mathbb{Z}}}^{'} \left[ i^2 + j^2 + k^2  \right]^{-\frac{s}{2}},
\end{aligned}
\end{equation}
\begin{equation} \label{eq:latticesum2}
\begin{aligned}
L_s^{\rm bcc}  &= {\sum\limits_{i,j,k \in \mathbb{Z}}}^{'} \left[ i^2 + j^2 + k^2 -\tfrac{2}{3}\left(ij +ik +jk\right) \right]^{-\frac{s}{2}},
\end{aligned}
\end{equation}
\begin{equation} \label{eq:latticesum3}
\begin{aligned}
L_s^{\rm fcc}  & = {\sum\limits_{i,j,k \in \mathbb{Z}}}^{'} \left[ i^2 + j^2 + k^2 +ij +ik +jk \right]^{-\frac{s}{2}},
\end{aligned}
\end{equation}
\begin{equation} \label{eq:latticesum4}
\begin{aligned}
L_s^{\rm hcp} & = {\sum\limits_{i,j,k \in \mathbb{Z}}}^{'} \left[ i^2 + j^2 + ij + \tfrac{8}{3}k^2\right]^{-\frac{s}{2}} \\
&+ \sum\limits_{i,j,k \in \mathbb{Z}}  \left[ \left( i + \tfrac{1}{3} \right)^2 + \left( j + \tfrac{1}{3} \right)^2 
+ \left( i + \tfrac{1}{3} \right)\left( j + \tfrac{1}{3} \right) + \tfrac{8}{3}\left( k + \tfrac{1}{2} \right)^2 \right]^{-\frac{s}{2}}.
\end{aligned}
\end{equation}
The notation $\sum^{'}$ implies that singularities in the sum at zero are avoided. Alternative decompositions to these expressions can also be found.\cite{burrows-2020}  A program to calculate these usually slow convergent lattice sums through various algorithms leading to fast converging series for real exponents $s\in\mathbb{R_+},s>3$ is freely available.\cite{ProgramJones} For example, for the (12,6) LJ potential we require the following values for lattice sums (given to 5 digits behind the decimal point): $L_6^{\rm sc}$=8.40192, $L_{12}^{\rm sc}$=6.20215, $L_6^{\rm bcc}$= 12.25367, $L_{12}^{\rm bcc}$= 9.11418, $L_6^{\rm fcc}$= 14.45392, $L_{12}^{\rm fcc}$=12.13188, and $L_6^{\rm hcp}$=14.45490, $L_{12}^{\rm hcp}$=12.13229.\cite{burrows-2020} We have for the limit ${\rm lim}_{s\rightarrow \infty} L_s=N_c$, where $N_c$ is the number of nearest neighbors in the crystal ($N_c$=6 for sc, 8 for bcc, and 12 for fcc and hcp), also called the kissing number. The lattice sums (minus the kissing number for better comparison) are shown in Figure \ref{fig:LatticeSums}.

\subsection{\label{sec:ELJ}Lattice vibrations for the cubic lattices}
As we move an atom in the crystal field of all other atoms, we break translational symmetry. Hence we need to apply a 3D Taylor expansion first to find appropriate formulae for the harmonic and anharmonic contributions to the total energy, and introduce the lattice sums, Eq.(\ref{eq:mapping}), in a subsequent step. Within the Einstein (E) model each atom of mass $M$ in the lattice is an independent 3D quantum harmonic oscillator,\cite{Einstein1907} i.e. all atoms oscillate with the same frequency $\omega_E$, whereas in the Debye model the atoms are assumed to be oscillating with their own frequencies and modes. For the zero-point vibrational energy contribution within the Einstein model we obtain a simple analytical formula for the three cubic lattices sc, bcc and fcc analogous to the simple harmonic oscillator formula (atomic units are used throughout),
\begin{equation} \label{eq:fccZPE}
E_{\rm ELJ}^{\rm ZPVE} = \frac{1}{2\sqrt{M}}\left( F_{xx}^{\frac{1}{2}} + F_{yy}^{\frac{1}{2}} + F_{zz}^{\frac{1}{2}}\right) = \frac{1}{2r_0}\sqrt{\frac{3}{M}}\left[ \sum_{n} s_n\left( s_n-1\right)c_n L_{s_n+2} r_0^{-s_n} \right]^{\frac{1}{2}}
\end{equation}
where the second derivative matrix $(F_{xy})$ denotes the harmonic force field. To obtain this expression, a selected atom is moved in an external ELJ field created by all the other atoms. The derivatives of the total energy with respect to the cartesian coordinates of a moving atom in a crystal lattice up to forth order, e.g. $F_{xyz...}=\partial^nE/\partial x\partial y\partial z...$, are detailed in the appendix. For the cubic lattices the Euclidean coordinate system $(x,y,z)$ is chosen parallel to the crystal axes such that $(F_{xy})$ is diagonal, and symmetry demands that $F_{xx}=F_{yy}=F_{zz}=tr(F)/3$. We mention that Corner also used a Taylor expression, but in his classical treatment for the vibrational movement, had to average over  the angular part.\cite{Corner-1939}

The ZPVE for the $(n,m)$ LJ potential, and more specifically, for the (12,6) LJ potential with coefficients $c_{1}=\epsilon r_e^{12}$ ($s_1=12$) and $c_2=-2\epsilon r_e^6$ ($s_2=6$) becomes (in atomic units),

\begin{equation}
\begin{aligned} \label{eq:EZPELJn-m}
E_{\rm LJ}^{\rm ZPVE}(r_{0}) &=
\frac{1}{2r_e}\sqrt{\frac{3\epsilon}{M}}\sqrt{\frac{nm}{n-m}}\left(\frac{r_e}{r_0}\right)^{n+1}\left[ (n-1) L_{n+2} -(m-1) L_{m+2} \left(\frac{r_0}{r_e}\right)^{n-m}\right]^{\frac{1}{2}}\\
&\quad\mathop{=}\limits_{m=6}^{n=12} \quad \frac{3}{r_e}\sqrt{\frac{\epsilon}{M}}\left(\frac{r_e}{r_0}\right)^{7}\left[ 11 L_{14} -5 L_8 \left(\frac{r_0}{r_e}\right)^{6} \right]^{\frac{1}{2}} \:,
\end{aligned}
\end{equation}
This expression is identical to that of Corner for a $(n,6)$-LJ potential.\cite{Corner-1939}
The (harmonic) Einstein frequency, $\omega_{\rm E}=2E_{\rm ELJ}^{\rm ZPVE}/3$, therefore becomes
\begin{equation} \label{eq:fccEinstein}
\omega_{\rm E} = \frac{1}{3\sqrt{M}}\left[3 \Tr\left(F\right) \right]^{\frac{1}{2}} = \frac{1}{r_0\sqrt{3M}}\left[ \sum_{n} s_n\left( s_n-1\right)c_n L_{s_n+2} r_0^{-s_n} \right]^{\frac{1}{2}} \,.
\end{equation}
The anharmonicity correction is usually small and can be obtained from first-order perturbation theory. Since the 3rd order term in the Taylor expansion around the origin is parity odd and the corresponding matrix elements thus equals zero, the anharmonicity correction is given by the corresponding matrix element of the 4th order term
\begin{equation} \label{eq:fccAZPVEME}
E^{\rm AZPVE} =  \frac{1}{24} \sum_{i=1}^{\infty}\sum_{n>3} c_n  \bra{\phi^{\rm E}_0}   \left( \vec{r}\cdot \vec{\nabla} \right)^4 |\vec{r}-\vec{r}_i|^{-n}\vert_{\vec{0}} \ket{\phi^{\rm E}_0},
\end{equation}
where the corresponding ground state harmonic oscillator (HO) solutions for a vibrating atom in 3D space is given by the Hartree product 
\begin{equation} \label{eq:fccEinsteinfunction}
\phi^{\rm E}_0 = \phi^{\rm HO}_0\left(x,\omega_{\rm E}\right) \phi^{\rm HO}_0\left(y,\omega_{\rm E}\right) \phi^{\rm HO}_0\left(z,\omega_{\rm E}\right) .
\end{equation}
This is very much in the spirit of the perturbative treatment for anharmonicity effects of a vibrating diatomic molecule. In first-order perturbation theory we only have to consider two matrix elements in the Taylor expansion for the ground vibrational state (apart from permutations in $x$, $y$ and $z$), $\left< \phi^{\rm HO}_0\left(x,\omega_{\rm E}\right) | x^2 | \phi^{\rm HO}_0\left(x,\omega_{\rm E}\right) \right>$ and $\left< \phi^{\rm HO}_0\left(x,\omega_{\rm E}\right) | x^4 | \phi^{\rm HO}_0\left(x,\omega_{\rm E}\right) \right>$, as all other quartic force constants with an odd number in one of the cartesian coordinates of the moving atom are zero due to crystal symmetry (and conveniently the cubic force field as well). The resulting anharmonic correction therefore becomes, 
\begin{equation} \label{eq:fccAZPVE}
E^{\rm AZPVE} = \frac{3}{32M^2\omega_E^2}\left( F_{xxxx} + 2F_{xxyy} \right).
\end{equation}
By using the results from the appendix we obtain for an ELJ potential,
%\begin{equation} \label{eq:fccAZPVE1}
%E_{\rm AZPVE} = \frac{1}{32M^2\omega_E^2} \sum_{i,n} \left( n+2 \right) \left( n+1 \right) n \left( n -1 \right) c_n r_i^{-n-4} 
%\end{equation}
%We now apply again the Lennard-Jones-Ingham procedure to obtain an analytical formula for the anharmonicity correction,
\begin{equation} \label{eq:fccAZPVE2}
E_{\rm ELJ}^{\rm AZPVE}(r_0) = \frac{1}{32M^2\omega_E^2} \sum_{n} \left( s_n+2 \right) \left( s_n+1 \right) s_n \left( s_n -1 \right) c_n L_{s_n+4} r_0^{-s_n-4} 
\end{equation}
and using Eq.(\ref{eq:fccEinstein}),
\begin{equation} \label{eq:fccAZPVE3}
E_{\rm ELJ}^{\rm AZPVE}(r_0) = \frac{3}{32Mr_0^2} \frac{\sum_{n} \left( s_n+2 \right) \left( s_n+1 \right) s_n \left( s_n -1 \right) c_n L_{s_n+4} r_0^{-s_n}} {\sum_{n} s_n\left( s_n-1\right)c_n L_{s_n+2} r_0^{-s_n}} .
\end{equation}
The AZPVE for the $(n,m)$ LJ potential, and more specifically, the (12,6) LJ potential with coefficients $c_{1}$  and $c_2$ as defined above becomes (in atomic units),
\begin{equation} \label{eq:AZPVELJ}
\begin{aligned}
E_{\rm LJ}^{\rm AZPVE}(r_{0}) 
& = \frac{3}{32Mr_0^2}  \frac{(n+2)(n+1)(n-1) L_{n+4} r_e^{n-m} - (m+2)(m+1)(m-1) L_{m+4} r_0^{n-m}} 
{(n-1) L_{n+2} r_e^{n-m} - (m-1) L_{m+2} r_0^{n-m}}  \\
& \mathop{=}\limits_{m=6}^{n=12}  \frac{21}{16M} \frac{20 L_{10} r_0^{6} - 143 L_{16} r_e^6} 
{5 L_{8}r_0^{8} - 11 L_{14}r_0^{2} r_e^6}  \:.
\end{aligned}
\end{equation}

This shows that by using the Einstein model, compact analytical expressions can be obtained for the vibrational contributions for the ELJ potential. Since the quartic force-constants are all positive, the anharmonicity correction {\it increases} the zero-point vibrational energy in contrast to a diatomic molecule, where a non-zero (negative) cubic force constant becomes important in second-order perturbation, leading to a decrease in the vibrational levels and transitions.

By defining the following sums,
\begin{equation} \label{eq:sums}
\begin{aligned}
A_L(r_0)= \: &r_0^{-2}\sum_{n} s_n\left( s_n-1\right)c_n L_{s_n+2} r_0^{-s_n},\\
B_L(r_0)= \: &r_0^{-2}\sum_{n} \left( s_n+2\right) s_n\left( s_n-1\right)c_n L_{s_n+2} r_0^{-s_n},\\
C_L(r_0)= \: &r_0^{-2}\sum_{n} \left( s_n+5\right) \left( s_n+2\right) s_n\left( s_n-1\right)c_n L_{s_n+2} r_0^{-s_n},\\
D_L(r_0)= \: &r_0^{-4}\sum_{n} \left( s_n+2 \right) \left( s_n+1 \right) s_n \left( s_n -1 \right) c_n L_{s_n+4} r_0^{-s_n},\\
%\end{aligned}
%\end{equation}
%\begin{equation} \label{eq:sums1}
%\begin{aligned}
%\nonumber
E_L(r_0)= \: &r_0^{-4}\sum_{n} \left( s_n+4\right) \left( s_n+2\right) \left( s_n+1\right) s_n\left( s_n-1\right)c_n L_{s_n+4} r_0^{-s_n} ,\\
F_L(r_0)= \: &r_0^{-4}\sum_{n} \left( s_n+7\right) \left( s_n+4\right) \left( s_n+2\right) \left( s_n+1\right) s_n\left( s_n-1\right)c_n L_{s_n+4} r_0^{-s_n} ,
\end{aligned}
\end{equation}
the volume/nearest neighbour distance expression for the ZPVE and anharmonicity corrections become,
\begin{equation} \label{eq:fccZPEvol}
E_{\rm ELJ}^{\rm ZPVE}(r_0) 
= \frac{1}{2}\sqrt{\frac{3}{M}} A_L(r_0)^{\frac{1}{2}}  
\end{equation}
and
\begin{equation} \label{eq:fccAZPVE3v}
\begin{aligned}
E_{\rm ELJ}^{\rm AZPVE}(r_0) 
=& \frac{3}{32M} A_L(r_0)^{-1} D_L(r_0) \:.
\end{aligned}
\end{equation}
Analytical expressions for the vibrational pressure and bulk modulus contributions for these cubic lattice can now be obtained. We get for the vibrational pressure,
\begin{equation} \label{eq:PZPVE}
\begin{aligned}
P_{\rm ELJ}^{\rm ZPVE}(r_0)
&=\frac{1}{4V\sqrt{3M}}A_L(r_0)^{-\frac{1}{2}} B_L(r_0) \:,
\end{aligned}
\end{equation}
\begin{equation} \label{eq:PAZPVE}
\begin{aligned}
P_{\rm ELJ}^{\rm AZPVE}(r_0)
=& \frac{1}{32VM} \left[ A_L(r_0)^{-1} E_L(r_0) - A_L(r_0)^{-2} B_L(r_0) D_L(r_0)\right] \:,
\end{aligned}
\end{equation}
and the bulk modulus,
\begin{equation} \label{eq:BZPVE}
\begin{aligned}
B_{\rm ELJ}^{\rm ZPVE}(r_0)
=& \frac{1}{24V\sqrt{3M}} A_L(r_0)^{-\frac{1}{2}} \left[2C_L(r_0) - A_L(r_0)^{-1} B_L(r_0)^2 \right] \:,
\end{aligned}
\end{equation}
\begin{equation} \label{eq:BAZPVE}
\begin{aligned}
B_{\rm ELJ}^{\rm AZPVE}(r_0)=& \frac{1}{96VM} A_L(r_0)^{-1} \big\{ F_L(r_0) - A_L(r_0)^{-1} \left[ 2B_L(r_0)E_L(r_0)+C_L(r_0)D_L(r_0)\right]  \\
&+2A_L(r_0)^{-2}D_L(r_0)B_L(r_0)^2 \big\} \:.
\end{aligned}
\end{equation}

%Gr{\"u}neisen parameter
\subsection{Gr{\"u}neisen parameter}
An important parameter in the theory of the equation of state and thermal expansion of solids is the volume (or pressure) and temperature dependent Gr{\"u}neisen parameter $\gamma(V,T)$, which describes the effect of changing the volume of a lattice on its vibrational properties.\cite{Barron1955,Poirier,Stacey2019,Anderson2000} At the microscopic level this parameter depends on the volume derivative of the phonon frequencies, and at $T=$0 K with wave vector $\vec{k}$ and band index $j$ the 
 dimensionless mode Gr{\"u}neisen parameter becomes,
\begin{equation} \label{eq:Grueneisendef}
\gamma_{\vec{k},j}(V)=-\frac{\partial{\rm ln}\omega_{\vec{k},j}(V)}{\partial {\rm ln}V} \:.
\end{equation}
For the Einstein approximation Eq.(\ref{eq:Grueneisendef}) simplifies to
\begin{equation} \label{eq:GrueneisenEinstein1}
\gamma_E(V)=-\frac{\partial{\rm ln}\omega_E(V)}{\partial {\rm ln}V},
\end{equation}
where we simply replaced the commonly used Debye frequency by the Einstein frequency. Using Eqs. (\ref{eq:sums}), (\ref{eq:fccZPEvol}) and (\ref{eq:PZPVE}) we obtain the ELJ potential
\begin{equation} \label{eq:GrueneisenEinstein2}
\gamma_{E, h}^{\rm ELJ}(r_0)=-\frac{V}{E_{\rm ELJ}^{\rm ZPVE}(r_0)}\frac{\partial E_{\rm ELJ}^{\rm ZPVE}(r_0)}{\partial V}=V\frac{P_{\rm ELJ}^{\rm ZPVE}(r_0)}{E_{\rm ELJ}^{\rm ZPVE}(r_0)}=\frac{B_L(r_0)}{6A_L(r_0)} \:.
\end{equation}
There is no mass dependence in $\gamma_E^{\rm ELJ}(r_0)$. The Gr{\"u}neisen parameter for the $(n,m)$ LJ potential, and more specifically, the (12,6) LJ potential with our coefficients $c_{1}$ and $c_2$ as defined above becomes (in atomic units),
\begin{equation} \label{eq:GrueneisenLJ1}
\begin{aligned}
\gamma_{E,h}^{\rm LJ}(r_0)&= \frac{1}{6}\frac{(n+2)(n-1)L_{n+2}\left(\displaystyle\frac{r_e}{r_0}\right)^{n-m}-(m+2)(m-1)L_{m+2}}{(n-1)L_{n+2}\left(\displaystyle\frac{r_e}{r_0}\right)^{n-m}-(m-1)L_{m+2}}\\
&\mathop{=}\limits_{m=6}^{n=12} \quad \frac{77L_{14}r_e^6-20L_8r_0^6}{33L_{14}r_e^6-15L_8r_0^6}
= \frac{77L_{14}V_e^2-20L_8V^2}{33L_{14}V_e^2-15L_8V^2}
\end{aligned}
\end{equation}
where $V_e$ is the volume at nearest neighbor distance $r_0=r_e$. The simplicity of this analytical formula demonstrates the beauty of the Einstein model.
In a similar way one can derive the anharmonicity contribution to the mode Gr{\"u}neisen parameter by the substitution $E_{\rm ELJ}^{\rm ZPVE}(r_0)\rightarrow E_{\rm ELJ}^{\rm ZPVE}(r_0)+E_{\rm ELJ}^{\rm AZPVE}(r_0)$, 
\begin{equation} \label{eq:GrueneisenEinstein3}
\gamma_{E, h+ah}^{\rm ELJ}(r_0)=V\frac{P_{\rm ELJ}^{\rm ZPVE}(r_0)+P_{\rm ELJ}^{\rm AZPVE}(r_0)}{E_{\rm ELJ}^{\rm ZPVE}(r_0)+E_{\rm ELJ}^{\rm ZPVE}(r_0)} \:,
\end{equation}
leading to a more complicated mass-dependent expression.

%The hexagonal close-packed structure
\subsection{\label{sec:hcp}The hexagonal close-packed structure}

Like fcc, the hcp lattice is a close-packed structure and often lies energetically very close to fcc. For the hard-sphere model the fcc and hcp packing densities are identical, as are any mixed fcc/hcp Barlow packings\cite{Barlow1883}. We remember that a cubic lattice is a lattice whose points lie at positions $(n_1,n_2,n_3)$ in the Cartesian three-space, where $n_i$ are integers. Unlike fcc however, the hcp lattice is not cubic and not a Bravais lattice, but instead belongs to the $D_{6h}$ point group. Although it has inversion symmetry, symmetry breaking occurs in the force field resulting in a lifting of the degeneracy of the Einstein frequencies. Hence, we lose the high symmetry compared to the three cubic lattices. This results in a far more complicated expression for the hcp compared to the fcc lattice sum, i.e. compare Eqs.(\ref{eq:latticesum1})-(\ref{eq:latticesum3}) with (\ref{eq:latticesum4}), which has been resolved in terms of fast converging series only very recently by our group\cite{burrows-2020}. 

The hcp lattice can be seen as a hexagonal Bravais lattice with lattice vectors $\vec{a}_1=\frac{a}{2}\hat{x}-\frac{\sqrt{3}a}{2}\hat{y}$, $\vec{a}_2=\frac{a}{2}\hat{x}+\frac{\sqrt{3}a}{2}\hat{y}$, $\vec{a}_3=c\hat{z}$, but with two atoms located at positions $\vec{r}_1^\top=(0,0,0)$ and $\vec{r}_2^\top=(2/3,1/3,1/2)$. Since each atom is experiencing exactly the same field from all other surrounding atoms in the bulk system, we only need to consider the summation over the many-body contributions from the atom placed at the origin for the cohesive energy. This implies that both atoms give the same diagonal 3D force field and the same set of Einstein frequencies. However, from the lattice vectors and the atom located at the origin it is clear that the vibration parallel to the hexagonal plane ($h$) axis will differ from the vibrations perpendicular to it ($c$). Thus we get for the diagonal force constants $F_{xx}=F_{yy}\ne F_{zz}$ and the corresponding three Einstein frequencies $\omega_1^h=\omega_2^h\ne\omega_3^c$. Even so we have relations between the different force constant for the hcp lattice as detailed by Wallace,\cite{Wallace1965} unfortunately for the Einstein frequency we have a sum of square-root terms for the force constants. Therefore the relations found for the cubic lattices cannot be applied anymore for the hcp structure.  Fortunately, it turns out that the difference $\Delta\omega=\omega_2^h-\omega_3^c$ is very small (of the order of 0.01 cm$^{-1}$ for argon) such that we can safely set $F_{xx}\approx F_{zz}$ and obtain to a very good approximation for hcp, this is the same expression in (\ref{eq:fccZPE}) for the ZPVE fcc case (with the corresponding hcp lattice sum). This holds for very small volumes (high pressures) as confirmed by numerical calculations carried out with our program SAMBA.\cite{ProgramSamba} The fact that Eq.(\ref{eq:fccZPE}) works is not surprising as we can use Corner's approximate treatment of vibrational motions applied to the hcp lattice.\cite{Corner-1939}

%now the cubic force field contributes (in second order)
Analyzing the higher derivatives for the hcp force field we obtain the symmetry relations for the quartic force constants $F_{xxxx}=F_{yyyy}\ne F_{zzzz}$ and $F_{xxzz}=F_{yyzz}\ne F_{xxyy}$ as discussed in detail by Wallace.\cite{Wallace1965} Again we see to a good approximation that $F_{xxxx}\approx F_{zzzz}$, but see larger differences for the mixed contributions in our numerical calculations. Fortunately, $F_{zzzz}\gg F_{xxzz}$ and therefore the AZPVE expression in Eq.(\ref{eq:fccAZPVE}) is applicable to a good approximation for the hcp lattice as well. For example, comparing both equations with numerical simulations for hcp argon at a volume set at 24 cm$^{3}/$mol (nearest neighbor distance of 3.8341 \AA~ close to the equilibrium distance), we obtain from numerical force field calculations the Einstein frequencies $\omega^h$=33.152 and $\omega^c$=33.141 cm$^{-1}$ and the ZPVE and AZPVE corrections of 49.7230 and 1.7758 cm$^{-1}$ respectively. This compares well with the ZPVE and AZPVE contributions from Eqs. (\ref{eq:fccZPE}) and (\ref{eq:fccAZPVE}) of 49.7230 and 1.7732 cm$^{-1}$ respectively, where the latter small difference could come from numerical inaccuracies.

%Temperature effects
\subsection{\label{sec:Temperature}Thermodynamics}
The thermodynamics of the solid state using the LJ potential has been reviewed by Anderson containing many useful formulae.\cite{anderson1989} The finite temperature contributions to the entropy and free energy may now also be expressed in terms of the lattice sums, using the expression for the Einstein frequency and the Boltzmann distribution. We start from the partition function for a single harmonic oscillator with frequency $\omega_i$,
\begin{equation} \label{eq:PartitionFunction}
Z_i = \frac{e^{-\beta \omega_i/2}}{1-e^{-\beta \omega_i}} \:,
\end{equation}
with $\beta=1/k_BT$, $T$ being the temperature and $k_B$ the Boltzmann constant converting the units of Kelvin to the desired energy unit. From this we get the phonon free energy for $N$ vibrating atoms, $F_{\rm vib} = -k_BT \ln Z$,
\begin{equation} \label{eq:PhononFreeEnergy}
F_{\rm vib} = \frac{1}{2}\sum_i^{3N}\omega_i + \beta^{-1}\sum_i^{3N} {\rm ln}\left( 1-e^{-\beta \omega_i}\right)
\end{equation}
which contains the zero-point vibrational contribution and the phonon entropy $ S = k_B T \partial \ln Z / \partial T + k_B\ln Z$,
\begin{equation} \label{eq:PhonoEntropy}
S_{\rm vib} = k_B\sum_i^{3N}\left[ {\rm -ln}\left( 1-e^{-\beta \omega_i}\right) 
- \frac{\beta \omega_i}{1-e^{-\beta \omega_i}} \right] \:.
\end{equation}
This expression trivially shows that for $T\rightarrow 0$ there is, aside the residual entropy, no entropy difference due to zero-point vibration between the lattices. For the Einstein approximation we obtain from Eq.(\ref{eq:PhononFreeEnergy}) the relation,
\begin{equation} \label{eq:PhononFreeEnergycubic}
F_{\rm vib} = \frac{3}{2}\omega_E + 3\beta^{-1}{\rm ln}\left( 1-e^{-\beta \omega_E}\right)
\end{equation}
and
\begin{equation} \label{eq:PhonoEntropycubic}
S_{\rm vib} = 3k_B\left[ {\rm -ln}\left( 1-e^{-\beta \omega_E}\right) 
- \frac{\beta \omega_E}{1-e^{-\beta \omega_E}} \right] \:.
\end{equation}
We obtain the following equation for the specific heat at constant volume ($F=E-TS$)
\begin{equation} \label{eq:SpecificHeat}
C_V=\left( \frac{\partial E}{\partial T} \right)_V = \frac{3}{4} k_B \left( \beta \omega_E \right)^2 \left[ \frac{e^{\beta \omega_E}}{\left( e^{\beta \omega_E}-1\right)^2} \right] \:.
\end{equation}

%Applications
\section{\label{sec:Applications}Applications}

In this section we apply our derived formulae for the LJ and ELJ potentials to the rare gas bulk phases of which the LJ potential already has a long history in the treatment of bulk systems.\cite{Pollack-1964,horton1976rare,Kalos1981,Schwerdtfeger-2006} Beside the simplicity of this model, for which we shall highlight the limitations, especially for a quantum system such as bulk helium, it offers qualitative yet valuable insight into bulk properties. Furthermore, these analytical formulae serve as a first good initial estimate of how important vibrational effects are for bulk quantities such as the equation of state. They also point towards further improvements like inclusion of higher body forces, phonon dispersion and, in the case of helium, dynamic effects to achieve better agreement with experimental observations. It should be borne in mind, however, that the rare gas solids represent a special case as the many-body expansion of the interaction energy converges reasonably well even at higher pressures.\cite{N-body.AH2007,Pahl_AngewChemIntEd47,SchwerdtfegerHermann2009,Schwerdtfeger-2016,Wiebke_AngewChemIntEd52}

\begin{table}[ht!]
\setlength{\tabcolsep}{0.16cm}
\caption{\label{tab:RareGases} (12,6) LJ  and ELJ properties for the fcc lattices of the rare gases at minimum energy. Binding energies -$\epsilon$, cohesive energies $E^{\rm stat}$, zero-point vibrational energies (ZPVE) $E^{\rm ZPVE}$ and anharmonicity corrections $E^{\rm AZPVE}$ in [$\mu$Ha] at $r_0^{\rm min}$. Equilibrium distances $r_e$ of the diatomic, nearest neighbour distance of the solid $r_0^{\rm min}$, ZPVE corrected nearest neighbour distance $r_0^{\rm ZPVE}$, critical distance $r_0^{\rm crit}$  and inflection point $r_0^{\rm infl}$ in [\AA]. Atomic masses $M$ used (in [amu]) are 3.016 and 4.003 for $^3$He and $^4$He respectively, 19.992 for $^{20}$Ne, 39.962 for $^{40}$Ar, 83.912 for $^{84}$Kr, 131.904 for $^{132}$Xe, 222.018 for $^{222}$Rn and 294.0 for $^{294}$Og. Binding energies and equilibrium distances for the (12,6) LJ potential are taken from Refs.\onlinecite{Cencek-2012,BichHellmannVogel2008,Jager_MolPhys107,Saue-2015}. The ELJ potential parameters are from: He (this work), Ne (Ref.\onlinecite{Hermann2009}), Ar (Ref.\onlinecite{Schwerdtfeger-2016}), Kr (Ref.\onlinecite{Smits2020}), Xe (Ref.\onlinecite{Smits2020}), Rn (Ref.\onlinecite{Smits2018}), Og (Ref.\onlinecite{Jerabek2019}).}
\begin{center}
\begin{tabular}{l|rrrrrrrrrr}
Isotope & \multicolumn{1}{c}{-$\epsilon$} & \multicolumn{1}{c}{$E^{\rm stat}$} & \multicolumn{1}{c}{$E^{\rm ZPVE}$} & \multicolumn{1}{c}{$E^{\rm AZPVE}$} & \multicolumn{1}{c}{$r_e$} & \multicolumn{1}{c}{$r^{\rm infl}$} & \multicolumn{1}{c}{$r_0^{\rm min}$} & \multicolumn{1}{c}{$r_0^{\rm ZPVE}$} & \multicolumn{1}{c}{$r_0^{\rm infl}$} & \multicolumn{1}{c}{$r_0^{\rm crit}$} \\
\hline
$LJ$\\                                                                                                                        
$^3$He	       	 & -34.8  	& -299.8  	& 462.3 	& 153.2	& 2.9676 	&  3.2901	& 2.8822 	& (3.3508)  	& 3.1955  & 3.3508 \\
$^4$He 		 & -34.8  	& -299.8  	& 401.3 	& 115.4	& 2.9676 	&  3.2901	& 2.8822 	& (3.3508)  	& 3.1955  & 3.3508\\
$^{20}$Ne 	 & -133.5 	& -1149.4 	& 337.7 	& 21.3 	& 3.0895 	&  3.4252	& 3.0006 	& 3.1250 		& 3.3267  & 3.4884\\
$^{40}$Ar 		 & -453.2 	& -3902.5	& 361.4 	& 7.2 	& 3.7618 	&  4.1706	& 3.6536 	& 3.6975 		& 4.0507  & 4.2476\\
$^{84}$Kr 		 & -636.1 	& -5477.3	& 276.8 	& 3.0		& 4.0158 	&  4.4523	& 3.9003 	& 3.9255 		& 4.3242  & 4.5344\\
$^{132}$Xe	 & -894.0 	& -7697.1	& 240.9 	& 1.6		& 4.3630  	&  4.8372	& 4.2375  	& 4.2543 		& 4.6980  & 4.9264\\
$^{222}$Rn	 & -1282.2& -11040	& 219.2 	& 0.9		& 4.4270  	&  4.9081	& 4.2997  	& 4.3104 		& 4.7670  & 4.9986\\
$^{294}$Og	 & -2844.3& -24490	& 290.1 	& 0.7		& 4.3290  	&  4.7995	& 4.2045  	& 4.2108 		& 4.6614  & 4.8880\\
\hline
$ELJ$\\
$^3$He		&  -34.9	& -258.1	& 432.0	& 113.9	& 2.9676	& 3.2906	& 2.9112	& (3.3530)		& 3.2322	& 3.3530\\
$^4$He 		&  -34.9	& -258.1	& 375.0	& 85.8	& 2.9676	& 3.2906	& 2.9112	& (3.3530)		& 3.2322	& 3.3530\\
$^{20}$Ne 	&  -132.2	& -1040.7	& 328.7	& 17.8	& 3.0930	& 3.4167	& 3.0278	& 3.1538		& 3.3501	& 3.4768\\
$^{40}$Ar 		&  -441.8	& -3470.0	& 346.7	& 5.5		& 3.7782	& 4.1731	& 3.7004	& 3.7430		& 4.0958	& 4.2460\\
$^{84}$Kr 		&  -636.1	& -4683.6	& 266.6	& 2.2		& 4.0157	& 4.4381	& 3.9346	& 3.9584		& 4.3577	& 4.5156\\
$^{132}$Xe	&  -894.3	& -6844.8	& 233.8	& 1.2		& 4.3616	& 4.8126	& 4.2782	& 4.2941		& 4.7309	& 4.8943\\
$^{222}$Rn	& -1212.9 	& -9665.3	& 202.9	& 0.6		& 4.4407	& 4.9153	& 4.3420	& 4.3520		& 4.8235	& 5.0004\\
$^{294}$Og	& -2853.6 	& -22482	& 258.1	& 0.4		& 4.3138	& 4.8259	& 4.1957	& 4.2011		& 4.7122	& 4.9256\\
\hline
\end{tabular}
\end{center}
\end{table}

\subsection{\label{sec:AppLJr}The equilibrium nearest neighbor distance and cohesive energy of the rare gas solids}
From the condition $ \partial E_{\rm ELJ}(r_0) / \partial r_0 = 0$ we derive the minimum nearest neighbor distance $r_0^{\rm min}$ of the atoms in the solid described by an ELJ potential. In the case of a general $(n,m)$ LJ potential we obtain a simple relationship between the equilibrium distance $r_e$ of the diatomic and the lattice $r_0^{\rm min}$ value,\cite{Schwerdtfeger-2006}
\begin{equation} \label{eq:LJdist}
 r_0^{\rm min}=\left( \frac{L_{n}}{L_m}\right)^{\tfrac{1}{n-m}}r_e \:.
\end{equation}
As for $n>m$ we have $L_n<L_m$ for a specific lattice,\cite{burrows-2020} we have $r_0^{\rm min}<r_e$. The same inequality holds for the ELJ potential as Table \ref{tab:RareGases} shows.

Using our analytical expressions we can determine the nearest neighbor distance for an ELJ potential including zero-point vibration. Table \ref{tab:RareGases} shows that vibrational effects increase the nearest neighbor distance in the solid, $r_0^{\rm ZPVE}>r_0^{\rm min}$, as pointed out earlier by accurate ab-initio calculations.\cite{Stoll-1999,Schwerdtfeger-2016,Smits2018,Jerabek2019,Smits2020} For example, the total cohesive energy for a (12,6) LJ potential including harmonic vibrational contributions within the Einstein approximation from Eqs. (\ref{eq:VolLJ}) and (\ref{eq:fccZPE}) becomes,
\begin{equation} \label{eq:TotalLJ} 
\begin{aligned}
E_{\rm LJ}^{\rm T}(r_0)&= E_{\rm LJ}^{\rm coh}(r_0) + E_{\rm LJ}^{\rm ZPVE}(r_0) \\
&= \frac{1}{2} \left( c_6 L_6 r_0^{-6} + c_{12} L_{12} r_0^{-12} \right)
+ \frac{3}{\sqrt{2M}}r_0^{-7}\left( 5c_6L_8 r_0^{6} + 22c_{12}L_{14} \right)^{\frac{1}{2}} \:.
\end{aligned}
\end{equation}
For the minimum $\partial E_{\rm LJ}^{\rm T}(r_0) /\partial r_0 = 0$ we get, after some algebraic manipulations, an 11th order polynomial in $x=r_0^2$,
\begin{equation} \label{eq:TotalLJexpr2}
a_0+a_3x^3+a_5x^5+a_6x^6+a_8x^8+a_9x^9+a_{11}x^{11} = 0
\end{equation}
with coefficients 
%$a_0=  88c_{12}^3L_{12}^2L_{14}$,
%$a_3=  4c_6c_{12}^2L_{12} \left( 22L_{6}L_{14} + 5L_{8}L_{12}\right)$,
%$a_5=  -11858M^{-1}c_{12}^2L_{14}^2$,
%$a_6=  2c_6^2c_{12}L_6\left( 11L_6L_{14} + 10L_8L_{12}\right),$,
%$a_8=  -3080M^{-1}c_6c_{12}L_{8}L_{14}$,
%$a_9=  5c_6^3L_6^2L_{8}$,
%$a_{11}= -200M^{-1}c_6^2L_{8}^2$.
%By substituting the LJ $c_6$ and $c_{12}$ parameters we can simplifying the formula,
$a_0=  44\epsilon L_{12}^2L_{14}$,
$a_3=  -4\epsilon r_e^{-6}L_{12} \left( 22L_{6}L_{14} + 5L_{8}L_{12}\right)$,
$a_5=  -5929M^{-1}r_e^{-12}L_{14}^2$,
$a_6=  4\epsilon r_e^{-12}L_6 \left( 11L_6L_{14} + 10L_8L_{12}\right)$,
$a_8=  3080M^{-1}r_e^{-18}L_{8}L_{14}$,
$a_9=  -20\epsilon r_e^{-18}L_6^2L_{8}$,
$a_{11}= -400M^{-1} r_e^{-24}L_{8}^2$,
the problem is then reduced to finding the zeros of the polynomial (\ref{eq:TotalLJexpr2}). There is no trivial solution except for $M\rightarrow \infty$, which yields just $r_0^{\rm min}$ for the minimum structure of the lattice and the polynomial has exactly one real solution. For a finite mass, the polynomial needs to be evaluated case by case. For all the rare gas solids, the polynomial has three real solutions and we find the second root to be the physical one. A similar expression can be obtained if the anharmonicity correction is added. 

\begin{figure}[t!]
\centering
\includegraphics[width=.48\columnwidth]{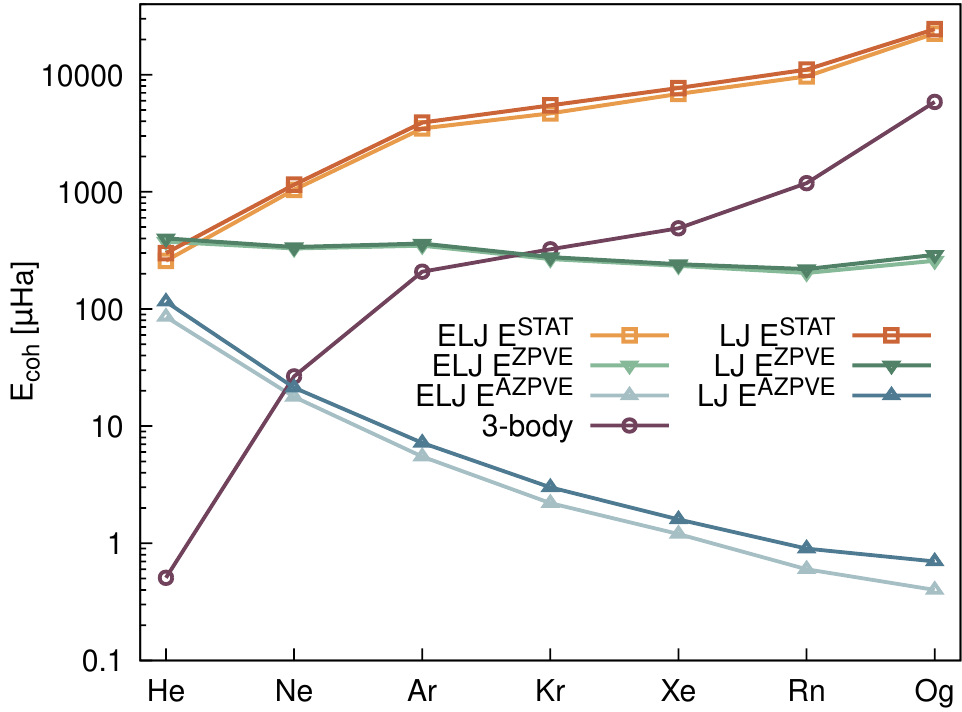}
\includegraphics[width=.48\columnwidth]{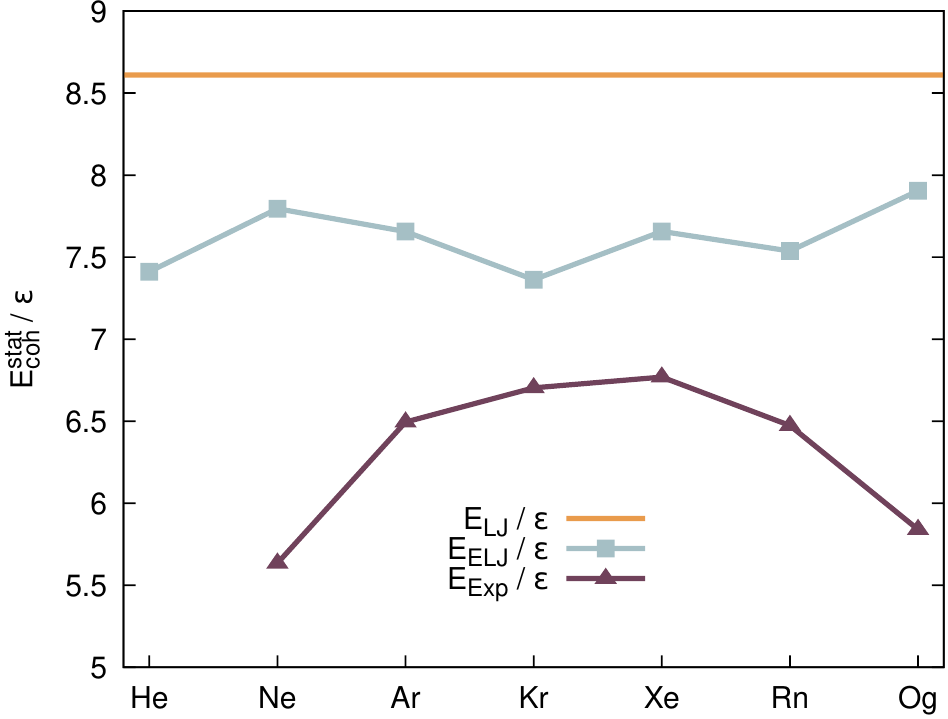}
\caption{
(a) Trends in cohesive energy contributions for $E^{\rm stat}$, $E^{\rm ZPVE}$, $E^{\rm AZPVE}$ and $E^{(3)}$ (in $\mu$Ha) shown at a logarithmic scale for all the rare gases. The values in Table \ref{tab:RareGases} were chosen, and for helium the $^4$He isotope was selected. The three-body contribution $E^{(3)}$ was taken from Ref.\onlinecite{Stoll-2000} for Ne to Xe, and Ref.\onlinecite{Jerabek2019} for Rn and Og. For He the program Samba was used and the three-body potential of Cencek, Patkowski and Szalewicz was taken (Ref.\onlinecite{Cencek-2009}) at the equilibrium distance $r_e$ for the dimer listed in Table \ref{tab:RareGases}. (b) Ratio between the two-body ELJ cohesive energy $E_{\rm ELJ}$ and the binding energy $\epsilon$ of the diatomic molecule (values taken from Table \ref{tab:RareGases}), and ratio for the best available cohesive energies\cite{McConville-1974,Schwalbe-1977,Stoll-1999,Stoll-2000,Jerabek2019} $E_{\rm coh}$ and $\epsilon$. The ideal LJ ratio is shown as a straight line.}
\label{fig:trend}
\end{figure}

Using Eq. (\ref{eq:LJdist}) for (\ref{eq:EcLJn-m}) we obtain a relationship for the cohesive energy at $r_0^{\rm min}$ in terms of the binding energy of the diatomic molecule and lattice sums,
\begin{equation} \label{eq:EcLJn-mmin}
E_{\rm LJ}(r_{0}^{\rm min})=\epsilon\frac{nm}{2(n-m)}\left[\frac{L_{n}}{n} \left(\frac{L_m}{L_n}\right)^\frac{n}{n-m} - \frac{L_m}{m} \left(\frac{L_m}{L_n}\right)^\frac{m}{n-m}\right] \quad\mathop{=}\limits_{m=6}^{n=12} \quad -\epsilon \frac{L_6^2}{2L_{12}} \:.
\end{equation}
For the fcc lattice we have $L_6^2/2L_{12}$=8.6102.\cite{Schwerdtfeger-2006} The ratios for the ELJ potential and with respect to the experimental or best theoretical values for the rare gas lattices are shown in Figure \ref{fig:trend}(b). There are two important messages we can deduce from this figure. First, the ELJ potential gives lower cohesive energies compared to the (12,6) LJ potential and the ratio $E_{\rm ELJ}/\epsilon$ varies slightly between 7.36 (Kr) and 7.90 (Og) compared to the LJ ideal value of 8.6102. Second, if we take the best available cohesive energy values for the rare gases to obtain the ratio $E_{\rm coh}/\epsilon$,\cite{McConville-1974,Schwalbe-1977,Stoll-1999,Stoll-2000,Jerabek2019} we see that zero-point vibrational effects lead to larger deviations for the lighter rare gas elements and the three-body effects to larger deviations for the heavier ones. 

\begin{figure}[t]
\centering
\includegraphics[width=.32\columnwidth]{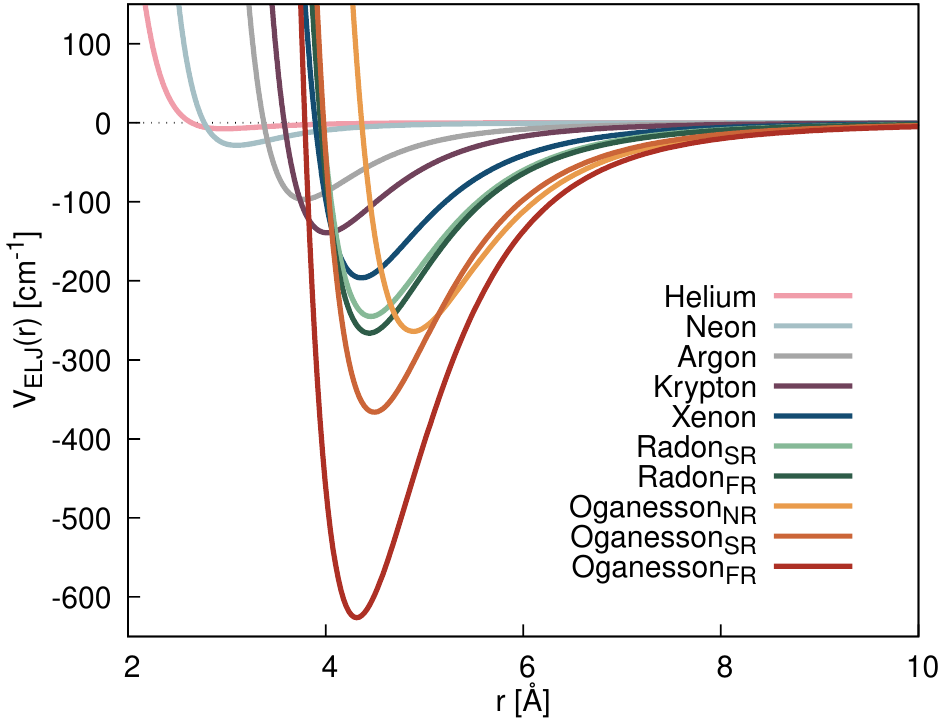}
\includegraphics[width=.32\columnwidth]{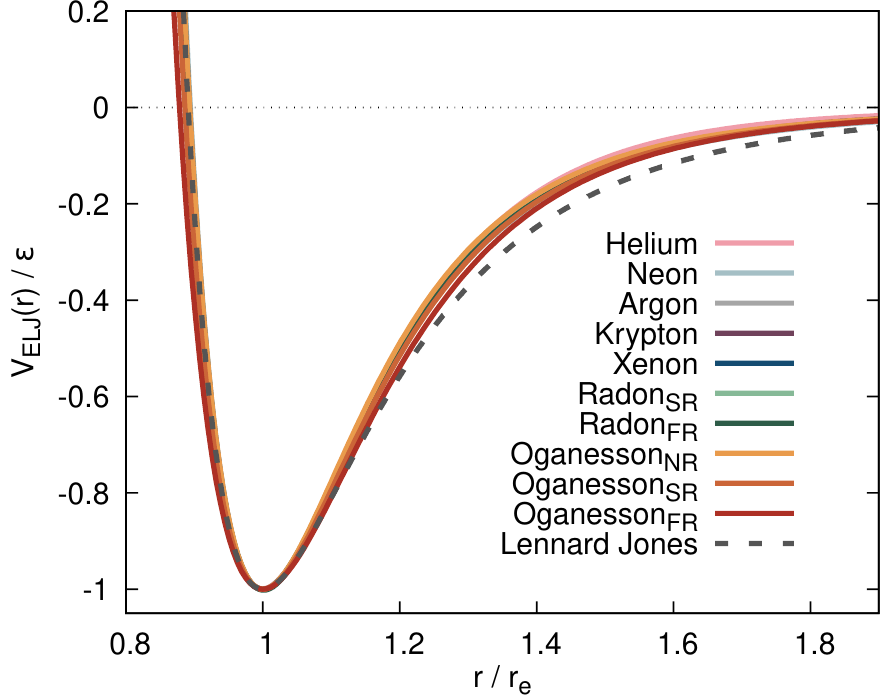}
\includegraphics[width=.32\columnwidth]{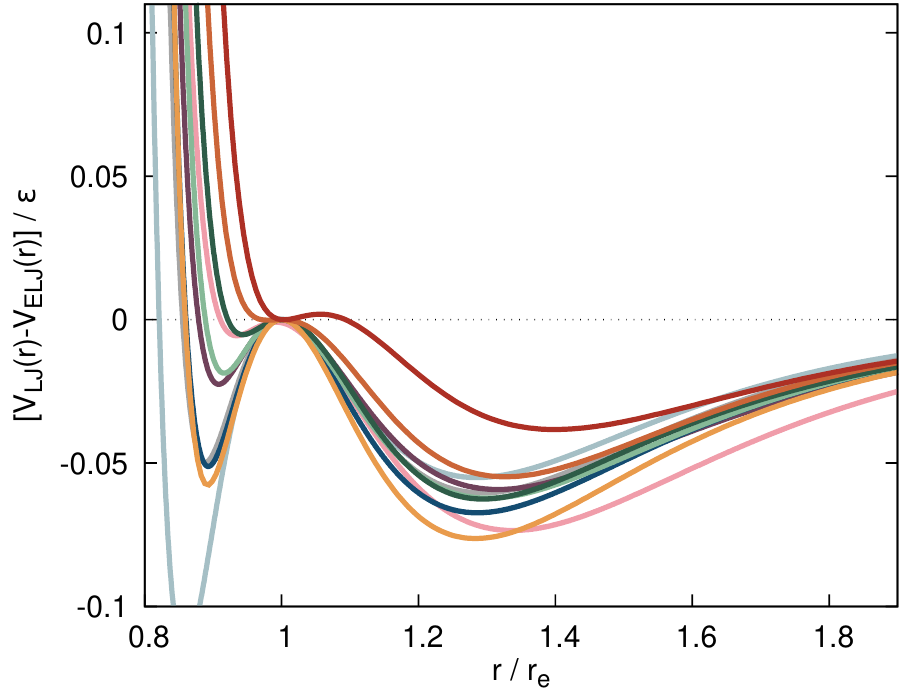}
\caption{(a) ELJ potentials of the noble gases, including potentials of Rn and Og at different levels of relativistic theory (NR: non-relativistic, SR: scalar relativistic, FR: fully relativistic (X2C). (b) All potentials rescaled to a potential with $r_e=1, \epsilon =1$. In grey the (12,6) LJ potential. (c) Difference between the unitary LJ and scaled ELJ potentials,  $\Delta V(r) = V_{\rm LJ}(r) - V_{\rm ELJ}(r)$  }
\label{fig:ELJcurves}
\end{figure}

Table \ref{tab:RareGases} shows properties for the fcc phase of the rare gas solids obtained by using both a (12,6) LJ as well as an ELJ potential with the values for the lattice sums $L_n$ published recently.\cite{burrows-2020} The corresponding potential curves are drawn in Figure \ref{fig:ELJcurves}(a) which show the very weak bonding for the lightest element, helium and the relatively strong bonding for the heaviest element in this group, oganesson. As can be seen from Figure \ref{fig:ELJcurves}(a), the unusually large cohesive energy of the heaviest known element in the periodic table is due to relativistic effects,\cite{Jerabek2019,Giuliani2019,schwerdtfeger2020a} which, despite the very large three-body contribution, results in a melting point above room temperature for oganesson.\cite{Schwerdtfeger2020} The trends in cohesive energy contributions for all the rare gas elements are summarized in Figure \ref{fig:trend}.

Concerning vibrational effects, we obtain a slow decrease in the ZPVE with increasing mass, gradually becoming less important compared to the static part of the cohesive energy. Oganesson is exceptional, since the increase in the cohesive energy and decrease in the bond distance, both due to relativistic effects, lead to a larger vibrational contribution compared to radon despite the larger mass. \cite{Jerabek2019} In contrast, anharmonicity effects diminish rather fast with increasing $Z$, see Figure \ref{fig:trend}. This can be understood from Eqs. (\ref{eq:fccZPE}) and (\ref{eq:fccAZPVE3}): For the ZPVE we have $E_{\rm LJ/ELJ}^{\rm ZPVE}\propto r_0^{-1}\sqrt{\epsilon/M}$. As $\epsilon$, $M$ and $r_0$ increase down the group in the periodic table we have a compensating effect and a small net decrease in the Einstein frequency. For the anharmonic contribution, however, we have $E_{\rm LJ/ELJ}^{\rm AZPVE}\propto r_0^{-2}M^{-1}$ leading to a much faster decrease in $E_{\rm LJ/ELJ}^{\rm ZPVE}$ with increasing mass and distance $r_0^{\rm min}$. 

To compare to experimental values we take solid argon as an example. The experimental nearest neighbor distance is 3.7560 \AA\cite{Barrett-1964} and the cohesive energy -2941(4) $\mu$Ha\cite{Schwalbe-1977}, in good agreement with the ELJ values of $E_{\rm ELJ}+E_{\rm ELJ}^{\rm ZPVE}+E_{\rm ELJ}^{\rm AZPVE}$= -3118 $\mu$Ha. If we take the optimized $r_0^{\rm AZPVE}$ distance instead, we obtain a similar value of 3134 $\mu$Ha), but the (12,6) LJ potential with -3534 $\mu$Ha clearly overestimates the cohesive energy. The remaining error for the ELJ potential lies mainly in the missing three-body effect. For a detailed analysis of the rare gas solids see Refs.\onlinecite{Stoll-1999,Schwerdtfeger-2016,Smits2018,Jerabek2019,Smits2020}. For comparison, we include three-body contribution from the literature in Figure \ref{fig:trend}, which shows that these effects become increasingly important with increasing nuclear charge and polarizability of the rare gas atom.\cite{AxilrodTeller1943}

Figure \ref{fig:ELJcurves}(b) compares the ELJ potentials by scaling both the equilibrium distance and the binding energy to unity. They all show a very similar functional form, the differences being barely visible on this graph. This suggests that, to a reasonable approximation, we can use the same analytical form for $\epsilon^{-1}E_{\rm ELJ}(r/r_e)$, which needs to be further investigated for the solid state properties of the rare gases. Figure \ref{fig:ELJcurves}(c) shows the difference between these curves and the standard (12,6) LJ potential. We see that the LJ potential over-binds in the long-range, but becomes too repulsive in the short-range, which will have consequences for the pressure-volume and bulk modulus-volume equation of states as we shall see below. However, before we proceed with the discussion of three of the rare gas solids, helium, neon and argon, we shall briefly discuss the analytical expressions for the critical points for the LJ and the ELJ potentials, and their relevance for the solid state.

%\subsection{\label{sec:CritPoint}Critical points for the extended Lennard-Jones potential energy curves}
\subsection{\label{sec:CritPoint}Critical points for the extended Lennard-Jones potential energy curves}
Multiple critical points, which are points on the function where the first or higher-order derivatives are equal to zero or a point where the function is not differentiable, for the ELJ potential can be identified. 

The first critical point is at the nearest neighbor distance $r_0^{\rm min}$ where the pressure is zero, $P = \partial E^{\rm coh}(r_0)/\partial r_0=0$. 
Expansion beyond the nearest neighbor distance into the region of negative pressure, $r>r^{\rm min}_0$,  is achieved by adding thermal pressure through the Boltzmann term, which keeps the pressure positive. The negative pressure range has been used, for example, to theoretically analyze the metal-to-nonmetal transition in expanded fluid mercury. \cite{Kresse1997}

A second critical point lies at the distance where $\partial^2 E^{\rm coh}(r_0)/\partial r_0^2=0$, referred to as the cohesive energy inflection point
\begin{equation} \label{eq:infl}
r_0^{\rm infl}=  \left[ \frac{(n+1)L_{n}}{(m+1)L_m}\right]^{\frac{1}{n-m}}r_e \:  
=  \left[ \frac{(n+1)}{(m+1)}\right]^{\frac{1}{n-m}}r_0^{\rm min} \:
 =  \left[ \frac{L_{n}}{L_m}\right]^{\frac{1}{n-m}}r^{\rm infl}  \:,
\end{equation} 
where $r^{\rm infl}$ is the inflection point of the (12,6) LJ  potential defined in Eq.(\ref{eq:LJ}). For the $(12,6)$ LJ potential we have $r_0^{\rm infl}= 1.07679 \:r_e = 1.10868 \: r_0^{\rm min}$. The restoring forces decrease with increasing deviations from equilibrium and at the inflection point the bulk modulus becomes zero, indicating that the compressibility becomes infinitely high, alike a gas at very low pressure.  Even though the lattice symmetry is maintained when moving along the cohesive energy curve, this hints that the inflection point can be used as a qualitative measure for symmetry breaking in the solid, resulting in a phase transition into the liquid or gas phase.\cite{cotterill1980melting,tallon1989}

Symmetry breaking occurs when one or more atoms in the lattice or unit cell move to positions where the lattice symmetry is not conserved, in contrast to expansion or compression of all atoms simultaneously of which the energy is given by the cohesive energy curve for the specific lattice symmetry. A good example for symmetry breaking is the so-called Peierls distortion (Jahn-Teller effect).\cite{peierls1996quantum,Kartoon2018} 

A local form of symmetry breaking happens when the Einstein frequency becomes zero and the square root in Eq. (\ref{eq:fccZPE}) or (\ref{eq:EZPELJn-m}) vanishes. This form of symmetry breaking was already discussed qualitatively for helium in 1955 by Houton \cite{Hooton1955,horton1976rare} and happens at a distance of 
\begin{equation} \label{eq:Rcrit}
r_0^{\rm crit} =
 \left[\frac{(n-1)}{(m-1)}\frac{L_{n+2} }{L_{m+2} }\right]^{\tfrac{1}{n-m}}r_e \: = \left[\frac{(n-1)}{(m-1)}\frac{L_{n+2} L_{m} }{L_{m+2} L_{n} }\right]^{\tfrac{1}{n-m}}r_0^{\rm min} \:.
\end{equation}
For the $(12,6)$ LJ potential, $r_0^{\rm crit}= 1.12912 \:r_e = 1.16257 \: r_0^{\rm min}$. Note that both Eq. (\ref{eq:infl}) and (\ref{eq:Rcrit}) are not mass dependent.
At expansion beyond $r_0^{\rm crit}$ a double minimum for the internal energy of the atom is formed, causing the atom to move away from the equilibrium distance and consequently the lattice locally distorts, breaking the symmetry of the bulk system. 
Yet, this simplified Einstein picture involves only the movement of one atom in the field of all other atoms which are kept at lattice symmetry points. If we allow all atoms in the solid to move, the point where symmetry breaks, $r_0^{\rm sb}$, lies below this Einstein estimate, $r_0^{\rm sb}<r_0^{\rm crit}$, and perhaps also below the inflection point for which we have $r_0^{\rm infl}<r_0^{\rm crit}$.

\begin{figure}
\centering
\includegraphics[width=0.45\columnwidth]{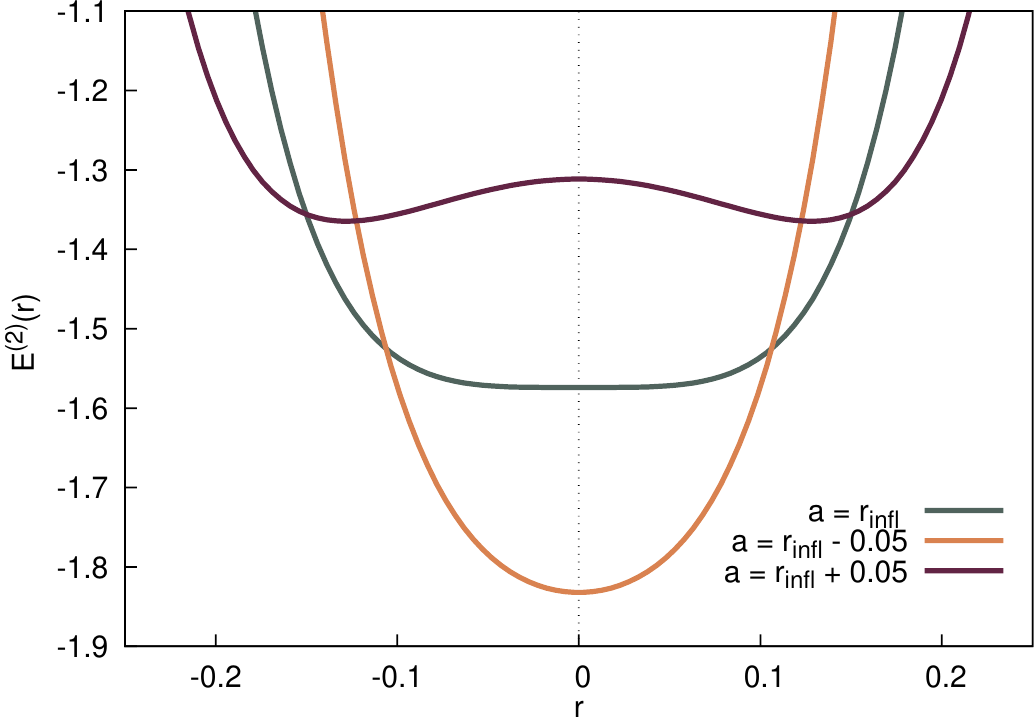}
\includegraphics[width=0.45\columnwidth]{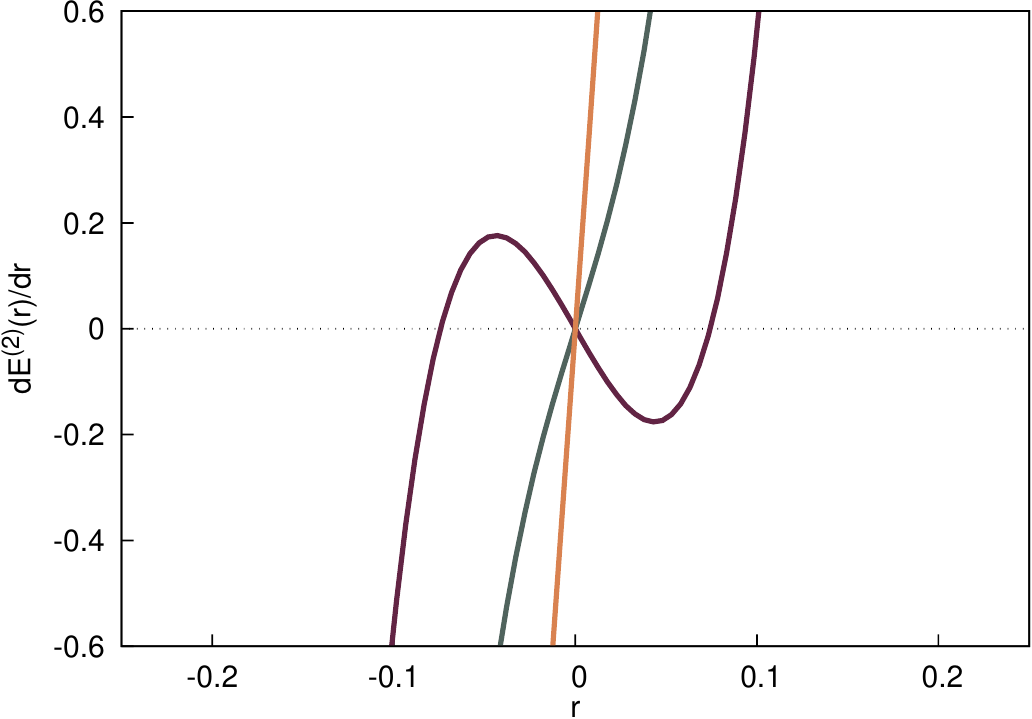}
\caption{LJ potential experienced by an atom confined by two other atoms to the left and right and separated by a distance of $2a$ resulting in a total interaction energy of  $E_{LJ}(r)=(r+a)^{-12}-2(r+a)^{-6}+(r-a)^{-12}-2(r-a)^{-6}$. The parameter used is (a) $a=r_{\rm infl}+0.05$,  (b) $a=r_{\rm infl}$, (c) $a= r_{\rm infl}-0.05$ (the inflection distance which is equal to the critical distance for a 1 dimensional chain). 
        (b) Corresponding effective on-site forces.}
\label{fig:symmetrybreaking}
\end{figure}

We briefly consider the inflection point and critical distance for close-packed structures in one and two dimensions for a LJ potential as they serve as a good models for symmetry breaking effects in solids. The expressions for the cohesive energy in Eq.(\ref{eq:EcLJn-m}), the inflection point, Eq.(\ref{eq:infl}), and critical distance, Eq (\ref{eq:Rcrit}), remain unchanged except that we have to substitute the 3D lattice sums $L_n^{\rm 3D}$ for the corresponding 1D or 2D ones. For a one-dimensional chain, these are related to the well-known Riemann zeta function, i.e. $L_n^{\rm 1D}=2\zeta(n)$ with the number of nearest neighbors $L_{\infty}^{\rm 1D}=2$, thus for the (12,6) LJ potential we have $\zeta(6)=\pi^6/945$, $\zeta(8)=\pi^8/9450$,  $\zeta(12)=691\pi^{12}/638512875$ and $\zeta(14)=2\pi^{14}/18243225$.  We obtain $r_0^{\rm 1D,infl}=1.10556 r_e$  and $r_0^{\rm 1D,crit}= 1.13967 r_e$. However, moving an atom in-between only two other atoms in one dimension, as shown in Figure \ref{fig:symmetrybreaking}, results in the equality $r_0^{\rm crit}=r^{\rm infl}$.

For the two-dimensional case the close-packed arrangement is the hexagonal lattice (one layer of the 3D fcc lattice) for which we can derive the corresponding lattice sums in terms of Riemann $\zeta(x)$ and Hurwitz $h(x,y)$ functions\cite{Coffey2008} according to Zucker and Robertson\cite{Zucker-1975a}, 
\begin{equation} \label{eq:zucker}
L_n^{\rm 2D}=3^{1-\tfrac{n}{2}}2\zeta(\tfrac{n}{2})\left[ h(\tfrac{n}{2},\tfrac{1}{3})- h(\tfrac{n}{2},\tfrac{2}{3})\right] \:.
\end{equation}
There are six nearest neighbors and therefore $L_{\infty}^{\rm 2D}=6$.\cite{,borwein1985,burrows-2020} 
We get $L_{6}^{\rm 2D}=6.37705$, $L_{12}^{\rm 2D}=6.01079$, $L_{8}^{\rm 2D}=6.10578$ and $L_{14}^{\rm 2D}=6.00382$.
This leads to $r_0^{\rm 2D,infl}= 1.09781 r_e$ and $r_0^{\rm 2D,crit}= 1.13724 r_e$.

For the fourth, and final, critical point let us discuss the minimal mass needed to stabilize the solid. 
Let us start with the minimal mass needed to form a bond between two atoms. Within the Born-Oppenheimer approximation two atoms can form a chemical bond if the ZPVE is smaller than the binding energy, $E_\mathrm{ZPVE}<\epsilon$. This implies that, within the harmonic approximation for the ground state vibrational energy level, we need 
\begin{equation} \label{eq:boundstate}
E_\mathrm{ZPVE}=\frac{6}{r_e}\sqrt{\frac{\epsilon}{M}}<\epsilon \:,
\end{equation}
from which we deduct the critical mass,
\begin{equation}
M_\mathrm{crit}^\mathrm{dimer}=\frac{36}{\epsilon r_e^2} \:.
\end{equation}
In this simple picture, $M>M_\mathrm{crit}^\mathrm{dimer}$ is thus required to stabilize a diatomic molecule E$_2$.  This is intuitive as a small binding energy requires a larger critical mass to stabilize a diatomic molecule within the Born-Oppenheimer approximation.  Using the values for helium in Table \ref{tab:RareGases} we obtain a critical mass of $M_\mathrm{crit}^\mathrm{dimer}= 17.9$ amu, which is far too high for any stable helium isotope. The harmonic ground state vibrational level lies above the diatomic He$_2$ potential curve,\cite{Lo_2008,Cencek2012} and only anharmonicity corrections, which are very large for this system due to the low mass and binding energy, together with an accurate treatment of the diatomic potential energy curve, can stabilize He$_2$ to such an extend that it can be observed at ultra-low temperatures.\cite{Fei1993,Schollkopf1994,Grisenti2000,Zeller2016} Yet, the remaining dissociation energy is very small for He$_2$,  \cite{Zeller2016} measured to be 5.58$\pm$0.49 nHa compared to the (uncorrected) binding energy shown in Table \ref{tab:RareGases}. In contrast, for Ne we obtain 4.3 amu  well below the mass of the most stable isotope of $^{20}$Ne.

\begin{figure}[h!]
\centering
\includegraphics[width=.49\columnwidth]{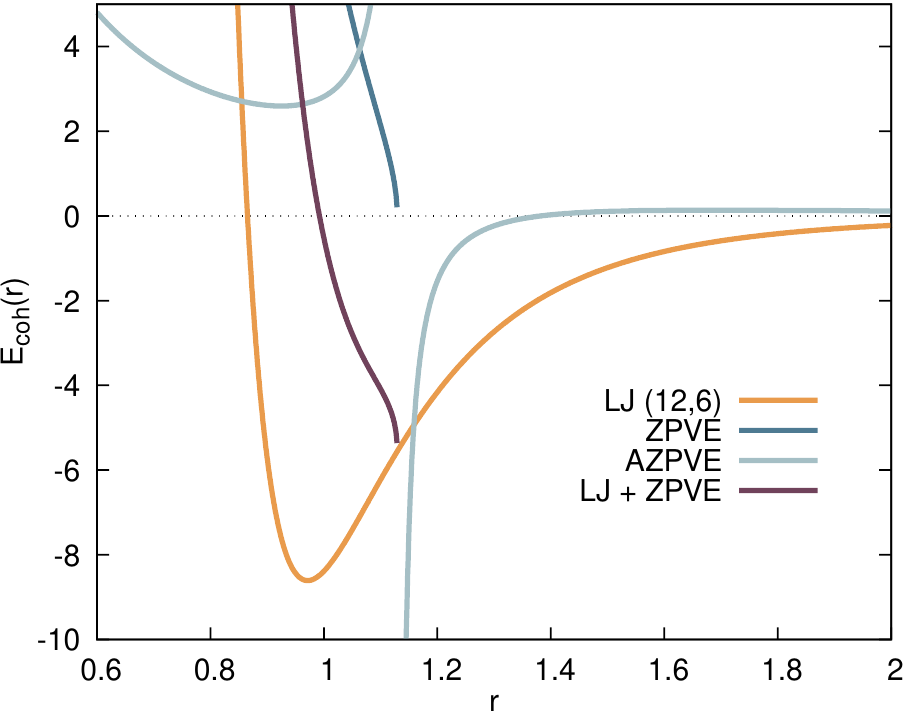}
\includegraphics[width=.49\columnwidth]{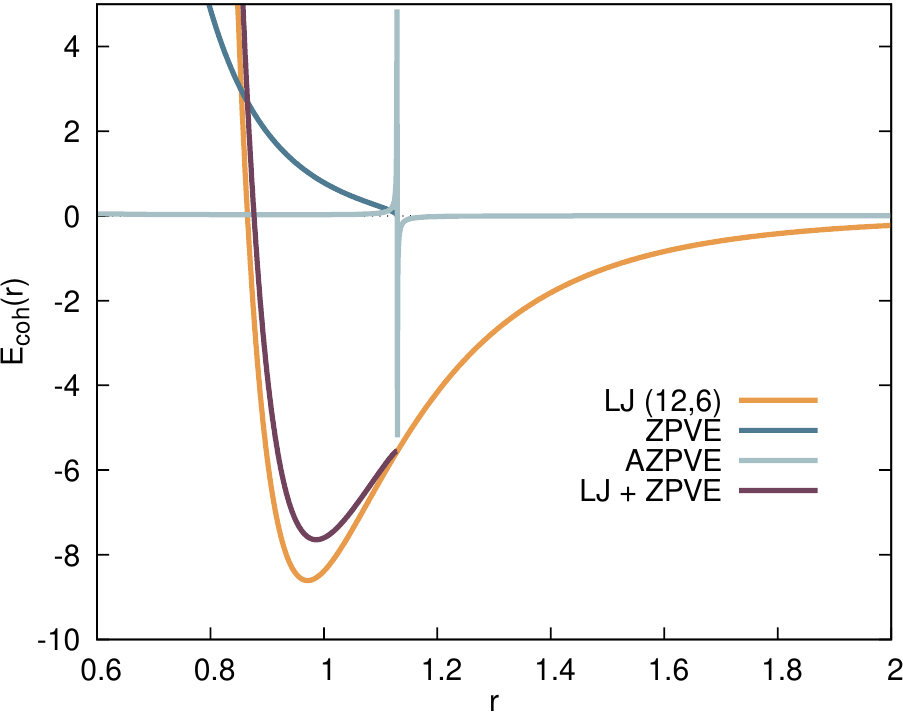}
\caption{(Color online) Static and dynamic contributions (only the real part of the ZPVE is shown, the expression for the ZPVE becomes complex beyond the critical distance) to the total cohesive energy for the (12,6) LJ potential ($\epsilon $ and $r_e$ set to unity), for the three different masses (a) $M = 10$ (b) $M = 100$ (c) $M = 1000$ according to Eq.(\ref{eq:boundstate}).}
\label{fig:CriticalMass}
\end{figure}

The same analysis may now be performed for the solid state, that is, we stabilize the solid described by a LJ potential if $E_{\rm LJ}^\mathrm{ZPVE}<-E_{\rm LJ}$ (remembering that $E_{\rm LJ}(r_0)$ was chosen to be negative in the attractive region).  Within the Einstein approximation we obtain the following relation from the combination of Eqs. (\ref{eq:EcLJn-m}), (\ref{eq:EZPELJn-m}) and (\ref{eq:LJdist}),
\begin{equation} 
M_\mathrm{crit}=\frac{36}{\epsilon r_e^2}f_\mathrm{solid} \quad \mathrm{with} \quad f_\mathrm{solid}=\frac{1}{L_6^3} 
\left(\frac{L_6}{L_{12}}\right)^{\tfrac{1}{3}} 
\left( 11L_{14}L_6 - 5L_8L_{12}\right)\:.
\end{equation}
This is identical to the result for the diatomic molecule except for the factor $f_\mathrm{solid}$. Using the lattice sums from Ref.\onlinecite{burrows-2020}, we get for the different structures $f_\mathrm{bcc}$=0.4298, $f_\mathrm{fcc}$=0.4005 and $f_\mathrm{hcp}$=0.4004. This reduces the helium critical mass to 7.17 amu for the fcc lattice compared to 17.9 amu for the diatomic.  However, the atomic critical mass is still too large for solid helium, i.e., the 8-He isotope has a half-life of 119 ms. Additionally, anharmonicity effects destabilize the rare gas solid. Phonon dispersion\cite{Eckert1977} most likely reduces the destabilizing harmonic ZPVE compared to the Einstein approximation,\cite{Schwerdtfeger-2016} and quantum effects beyond the Born-Oppenheimer approximation also become important for the treatment of solid helium.\cite{Ceperley1995} 

Figure \ref{fig:CriticalMass} shows the cohesive energy of the (12,6) LJ potential with $M$ below and with mass $M$ above the critical mass respectively. If the mass $M$ is small, as it is for $^3$He or $^4$He, the vibrating periodic lattice does not have a minimum, see Figure \ref{fig:CriticalMass}(a) where the potential curve for $E_{\rm LJ}(r)+E_{\rm LJ}^{\rm ZPVE}(r)$ abruptly ends when $\omega_{\rm E}$ becomes imaginary.  Hence, the $r_0^{\rm ZPVE}$ values for helium are set in parentheses as this is the point when the lattice optimization stops because of $\omega_E=0$. Here the perturbative treatment for anharmonicity effects completely breaks down. At larger masses the minimum is retained, see Figure \ref{fig:CriticalMass}(b).

Experimentally, it is known that under pressures of approximately 2.5 MPa helium is quite unusual as it solidifies to the hcp phase,\cite{simon1950liquid,Ceperley2012} and a hcp$\rightarrow$fcc phase transition occurs at 1.1 GPa and 15K.\cite{Dugdale1953} Helium at extreme conditions plays an important role within the science of planets and stars. \cite{Saumon_2004,Stixrude2008,Ceperley2012,Khairallah2008} We therefore discuss the validity of the (12,6) LJ model for the less critical helium high-pressure range in the following section.

%The extended Lennard-Jones equation of state for solid helium
\subsection{\label{sec:Helium}The equation of state for solid helium}

\begin{figure}[h!]
\centering
\includegraphics[width=.48\columnwidth]{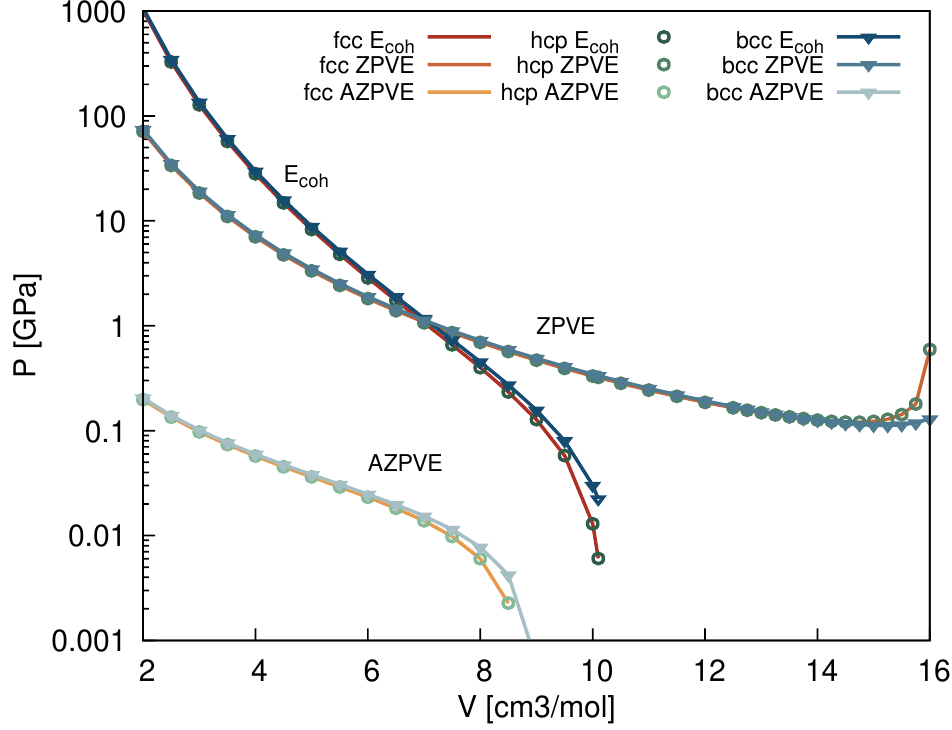}
\includegraphics[width=.48\columnwidth]{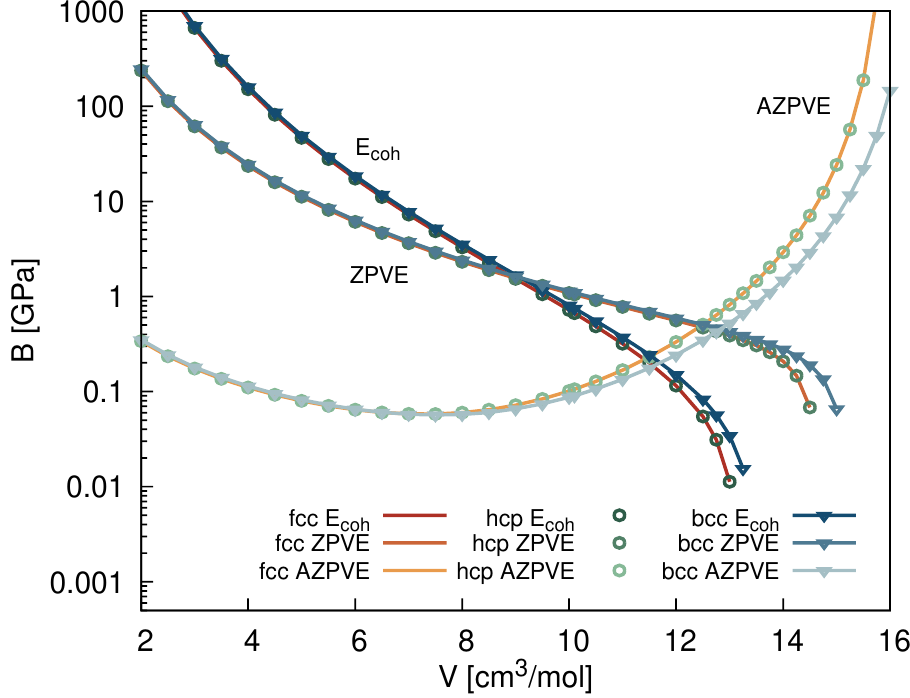}
\includegraphics[width=.48\columnwidth]{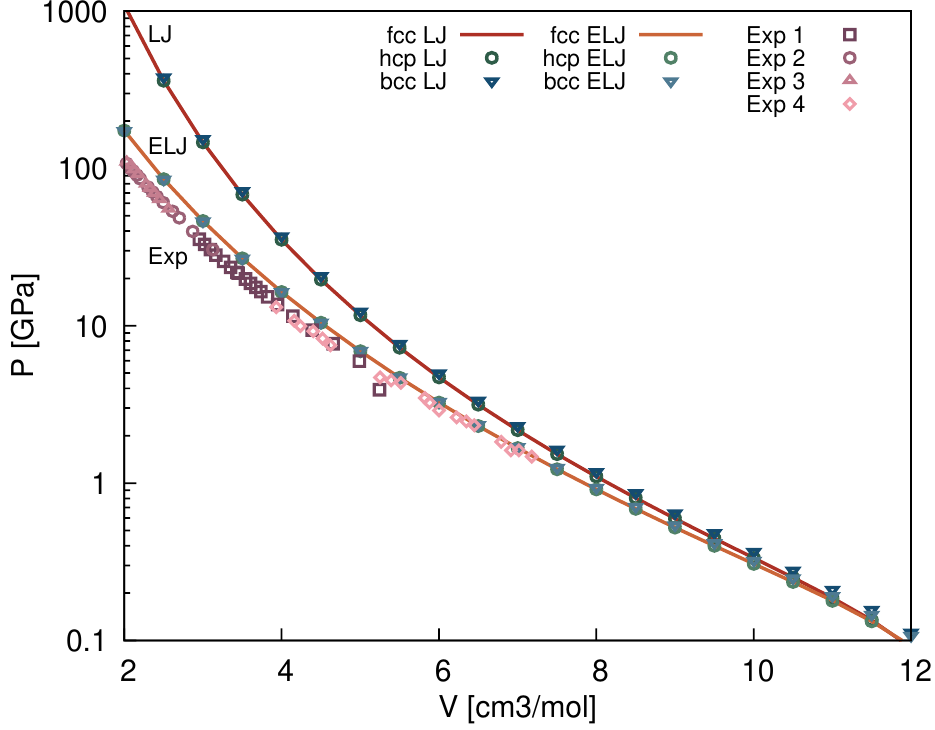}
\includegraphics[width=.48\columnwidth]{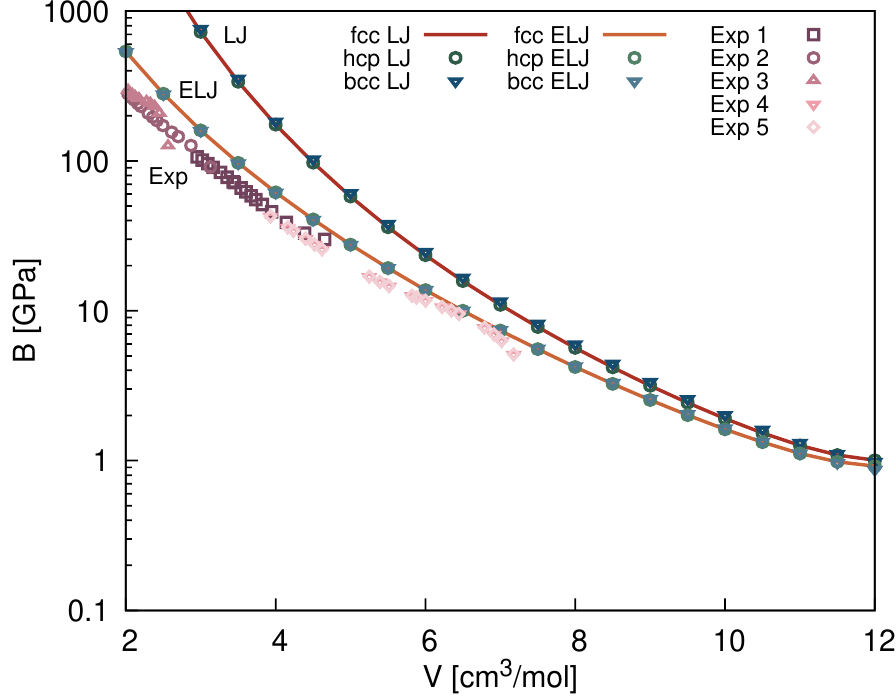}
\caption{(Color online) Pressure $P(V)$ and bulk modulus $B(V)$ curves for the fcc, hcp and bcc phases of solid helium derived from the analytical formulae presented in this paper (logarithmic scale is used for $P$ and $B$). (a) (12-6) LJ $P(V)$-diagram for the different pressure contributions to the static cohesive energies $P_{\rm LJ}$, harmonic zero-point vibrational $P_{\rm LJ}^\mathrm{ZPVE}$ and anharmonic contributions $P_{\rm LJ}^\mathrm{AZPVE}$ within the Einstein approximation. (b) Same as (a) but for the bulk modulus $B(V)$. (c) Total pressure $P=P_{\rm ELJ}^\mathrm{stat}+P_{\rm ELJ}^\mathrm{ZPVE}+P_{\rm ELJ}^\mathrm{AZPVE}$ for the LJ and ELJ potentials in comparison to experimental data from Refs.\onlinecite{Dawaele2020,Hemley1988,Hemley1993}. Exp1: $T$=15K, pressure gauge (PG): SrB$_4$O$_7$:Sm$^{2+}$; Exp2: $T$=297K, PG: W; Exp3: $T$=297K, PG: ruby; Exp4: $T$=300K, PG: ruby. (d) Exp1-Exp3 as (c) for the bulk modulus $B(V)$. Exp5: extrapolated to $T$=0K , isochor cell, from Ref.\onlinecite{Driessen1986}. For the conversion of pressure units we used 1 a.u. = 2.94210157$\times 10^4$ GPa. For hcp we took the ideal $c/a=\sqrt{8/3}$ ratio as lattice distortions are small even at higher pressures.\cite{Freiman2009}}
\label{fig:LJnm}
\end{figure}

Figure \ref{fig:LJnm} shows LJ $P(V)$- and $B(V)$ curves for solid helium for the three different observed phases fcc, hcp and bcc of $^4$He in the pressure/volume range where this simple LJ Einstein model should work reasonably well.\cite{Edwards1965,Edwards1966,Holian1973,McMahan1981} To give a feeling for the volume range to be considered for bulk helium, the liquid state of $^4$He at normal pressure has a density of 0.125 g/cm$^3$ corresponding to a very large volume of 32 cm$^3$/mol.\cite{lide2004crc} In contrast, solid helium has a density of 0.214 g/cm$^3$ at 6.7 GPa  corresponding to a volume of 18.7 cm$^3$/mol and nearest neighbor distance of $r_0$=3.528 \AA, which is larger then both the inflection point, $r_0^{\rm infl}$, and critical distance, $r_0^{\rm crit}$, see Table \ref{tab:RareGases}. This shows the limitation of the simple Einstein model for bulk helium.\cite{Henshaw1958} Indeed, in this very low density range, zero-point vibrational energy effects dominate for both the pressure and the bulk modulus as can be seen from Figures \ref{fig:LJnm}a and \ref{fig:LJnm}b. 

Solid helium shows giant plasticity and superfluid-like mass transport at large volumes and low temperatures\cite{Hunt2009,Borda2016} (for a recent review see Beamish and Balibar\cite{Beamish2020}), and our 'static' model used here cannot accurately describe such phenomena. Moreover, at these large volumes perturbation theory used for the anharmonicity effects breaks down and one requires a full dynamic treatment, for example by using quantum Monte-Carlo simulations.\cite{Ceperley1995,Ceperley2012,Cazorla2015} This can already be seen for the bulk moduli at volumes $V>12$ cm$^3$/mol, where the LJ results start to deviate substantially from the experimental results, see Figure \ref{fig:LJnm}d. We therefore focus on the high pressure regime instead.

Gr\"uneisen already pointed out in 1912 that the vibrational frequency increases with pressure\cite{Gruneisen1912} because the potential energy becomes increasingly repulsive. Our Einstein model shows that harmonic vibrational contributions to the pressure dominate down to volumes of 8 cm$^3$/mol.  Below 8 cm$^3$/mol, the pressure contribution coming directly from the static cohesive energy (\ref{eq:PLJ}) starts to dominate over vibrational effects. A similar behavior is observed for the bulk modulus. Here, anharmonicity effects become even more important in the low density range. As helium represents a special case within the rare gas elements, \cite{horton1976rare} for the heavier rare gases this picture changes significantly because of the increasing mass.\cite{Stoll-1999,Stoll-2000,SchwerdtfegerHermann2009}

We can determine the point at which the vibrational pressure becomes less important than the pressure created by the repulsive wall of the potential energy curve for a LJ potential, that is $P_{\rm LJ}^{\rm ZPVE}(V)=P_{\rm LJ}(V)$ at a specific volume, which we denote as $V_\mathrm{H}$. For a (12,6) LJ potential we get a simple relation from Eqs.  (\ref{eq:PLJ}) and (\ref{eq:PZPVE}),
\begin{equation} \label{eq:Pvib}
f\left( V_\mathrm{H} / V_e \right)=\epsilon r_e^2 M \:,
\end{equation}
where $f(x)$ is an algebraic function containing only the lattice sums for a specific lattice,
\begin{equation} \label{eq:Pvib1}
f(x)= {\left( \frac{L_6}{L_{12}} \right)}^{\frac{1}{3}}  \frac{ x^4 \left( 77L_{14}-20L_8x^2\right)^2}{4\left( L_{12}-L_6x^2\right)^2\left( 11 L_{14}-5L_8x^2  \right)} \:.
\end{equation}
The left- and right-hand side of Eq. (\ref{eq:Pvib}) are dimensionless (either use atomic units for calculating $\epsilon r_e^2 M$ or divide this expression by $\hbar^2$). As the pressure from the cohesive energy is zero at the minimum distance, the validity range is $x=V_\mathrm{H}/V_e\ll 1$. In any case, from the data in Table \ref{tab:RareGases} we get $\epsilon r_e^2 M=8.048$ for $^4$He and $\epsilon r_e^2 M=6.065$ for $^3$He corresponding to a volume ratio of $V_\mathrm{H}/V_e=0.647$ and $V_\mathrm{H}/V_e=0.624$ for the fcc lattice respectively (for comparison for $^{20}$Ne we have $V_\mathrm{H}/V_e=0.829$ and for $^{40}$Ar 0.886, much closer to the minimum value $V=V^{\rm min}/V_e=(r_0^{\rm min}/r_e)^3=0.916$). This demonstrates the importance of vibrational effects for $^4$He and $^3$He in the low to medium pressure range because of their low mass.

The question now rises how well this (12,6) LJ model works. As already mentioned, in the low density range one requires a more complete quantum picture not considered here.\cite{Eckert1977,Ceperley1987,Ceperley1995,Ceperley2012} In the high density range we can compare to experimental data from Dawaele\cite{Dawaele2020} as shown in Figure \ref{fig:LJnm}(c) and (d) (when bulk experimental moduli were not available, a polynomial fit to the observed $P(V)$ data was used to obtain $B(V)$). The data show that the LJ $P(V)$ curve (containing all terms within the Einstein approximation) deviates substantially from the experimentally obtained values,\cite{Dawaele2020} and increasingly so with decreasing volume. These large deviations in the high pressure range are mostly due to the incorrect repulsive form of the (12,6) LJ potential as has been pointed out before.\cite{Polian1986,Loubeyre1987}  

More accurate two-body potentials $V^{(2)}(r)$ are known for all the rare gases up to oganesson,\cite{Kalos1981,Ross1986,Aziz1991,BichHellmannVogel2008,Cencek-2012,Jager_MolPhys107,Saue-2015,Przybytek2017} and there are already a number of theoretical studies for the $P(V)$ curves of solid helium.\cite{Moroni2000,Herrero_2006} To further investigate the failure of the LJ potential in the high pressure range we fitted a recently published potential energy curve $V^{(2)}_\mathrm{PCJS}$ by Przybytek, Cencek, Jeziorski, and Szalewicz (PCJS) for the helium dimer,\cite{Przybytek2017} who included adiabatic, relativistic as well as QED effects in their coupled-cluster treatment, to an ELJ potential. We used a least-squares fit procedure introducing distance dependent weights $\omega(r)$ to take care of the very small and large energy values in the long- and short-range of the potential energy curve respectively,
\begin{equation} \label{eq:LeastSquare}
\frac{\partial}{\partial c_m} \int_{r_{\rm c}}^{\infty}dr ~\omega(r)\left[ \sum_{n=1}^N c_nr^{-s_n} - V^{(2)}_\mathrm{CS}(r)\right]^2 = 0 \:,
\end{equation}
which leads to a set of $N$ linear equations for the coefficients $c_n$ ($m=1,...,N$),
\begin{equation} \label{eq:LeastSquare}
\sum_{n=1}^N c_n \int_{r_{\rm c}}^{\infty}dr ~\omega(r)  r^{-(s_n+s_m)} = \int_{r_{\rm c}}^{\infty}dr ~\omega(r) r^{-s_m} V^{(2)}_\mathrm{PCJS}(r) \:.
\end{equation}
 We applied a numerical integration scheme, a weighting function of $\omega(r)=1-e^{-ar}$ with $a=0.89$, and set $r_{\rm c}$ to the lowest possible value of 2.1 a.u. to obtain a good fit over the whole distance range. The resulting ELJ potential yields an equilibrium distance of $r_e=$ 5.6080 a.u., an inflection point $r^{\rm infl}$ at  6.2183 a.u. and a binding energy of $\epsilon=$-348.746 $\mu$Ha compared to the PCJS potential of 5.6080 a.u,  6.2089 a.u. and -348.236 $\mu$Ha respectively. This should give accurate two-body pressures up to about 1 TPa. We fixed the parameter $c_1=-C_6$ and $c_2=-C_8$ to the Van der Waals coefficients given in Ref.\onlinecite{Przybytek2017} to correctly describe the long-range, and chose $c_{N}>0$ to correctly describe the repulsive short-range. The obtained parameters $c_n$ are listed in Table \ref{tab:2bodyfitting}. 
\begin{table}[ht!]
\centering
\caption{Potential parameters for the He dimer obtained from a least-squares fit to the analytical form of Szalewicz and co-workers\cite{Przybytek2017}. All potential parameters are given in atomic units.}
\label{tab:2bodyfitting}
\begin{tabular}{r|l|r|r|l|r}
 \hline  \hline
 $n$ & $s_n$  & $c_n$ & $n$ & $s_n$  & $c_n$   \\
\hline
1  	& 6   & -1.4618550565137 &  2 & 8 & -14.1208183897247  \\
3  	& 9   & 13997.975339736 &  4 & 10 & -304327.625470953  \\
5  	& 11 & 2441586.03190761 &  6 & 12 & -8163337.07262287  \\
7  	& 13 & 3390456.21241699 &  7 & 14 & 51324186.4628455  \\
9  	& 15 & -118039510.368528 &  10 & 16 & -31496186.3299036  \\
11 	& 17 & 456234485.18761 &  12 & 18 & -639488529.764361  \\
13 	& 19 & 296722948.860609 &   &  &  \\
\hline  \hline
\end{tabular}
\end{table}

\begin{figure}[h!]
\centering
\includegraphics[width=.5\columnwidth]{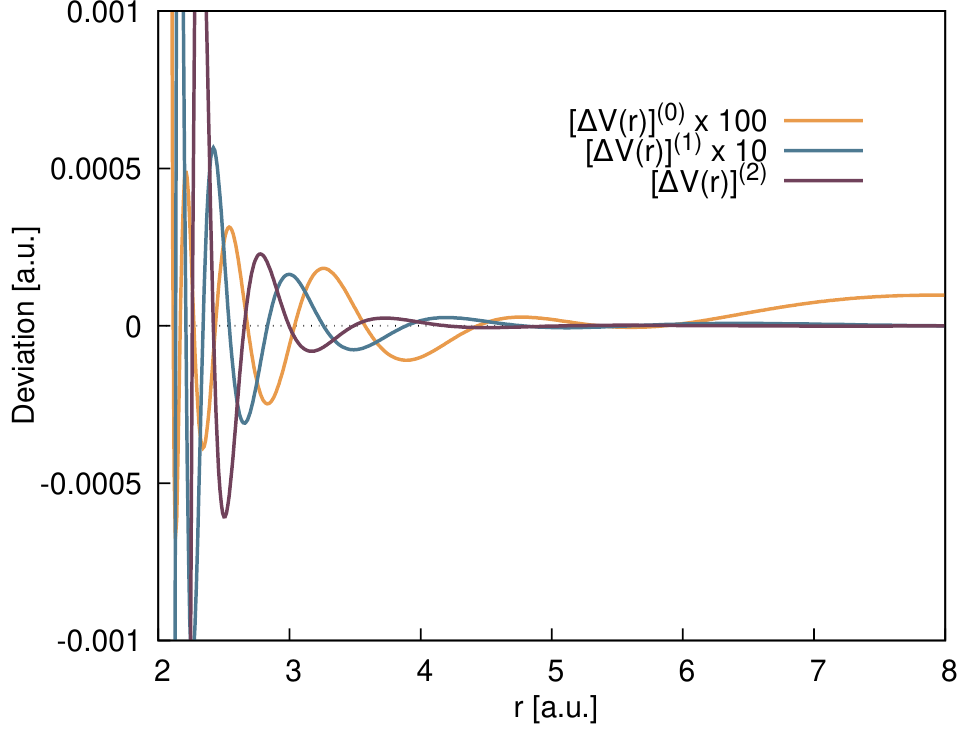}
\caption{(Color online) Deviations between the ELJ and the PCJS potential, $[V^{(2)}_\mathrm{ELJ}(r)-V^{(2)}_\mathrm{PCJS}(r)]^{(n)}$, up to second order in the derivatives ($n\le2$).}
\label{fig:fderiv}
\end{figure}

Figure \ref{fig:fderiv} shows the deviations $[V^{(2)}_\mathrm{ELJ}(r)-V^{(2)}_\mathrm{PCJS}(r)]^{(n)}$ up to the second derivatives ($n=2$). As can be seen, the error in the energy is of the order of a few $\mu$Ha which is acceptable and the error in the first and second derivatives increase by an order of magnitude each. A test calculation with our program SAMBA\cite{ProgramSamba} ensured that in the distance range $r>2.1$ a.u. ($V>0.6$ cm$^3$/mol for the fcc structure) the energy, pressure and bulk moduli are in very good agreement with the results from the PCJS potential. For example, the two-body cohesive energy, pressure and bulk modulus at $V=1$ cm$^3$/mol for the ELJ and  PCJS potential (the latter obtained numerically and given in parentheses) are $P=$1.2187 (1.2183) TPa and $B=$3.221 (3.227) TPa.

To compare with experimental $P(V,T)$ and $B(V,T)$ data, one has to include the increase in pressure and bulk modulus due to finite temperature effects. For this we use the Einstein approximation (\ref{eq:PhononFreeEnergycubic}) to obtain the thermal phonon pressure ($\beta=1/k_BT$),
\begin{equation} \label{eq:PfiniteT}
P_{th}\left(V,T\right)=-3\frac{\partial\omega_E(V)}{\partial V}\left[e^{\beta\omega_E(V)} -1\right]^{-1}=2P_{\rm ZPVE}\left[e^{\beta\omega_E(V)} -1\right]^{-1}
\end{equation}
and similar for the bulk modulus,
\begin{align} \label{eq:BfiniteT}
B_{th}\left(V,T\right)&= B_{th1}\left(V,T\right)+B_{th2}\left(V,T\right)\\ \nonumber
&=2B_{\rm ZPVE}\left(V\right)\left[e^{\beta\omega_E(V)} -1\right]^{-1}-\frac{4}{3}P_{\rm ZPVE}^2\left(V\right)\beta Ve^{\beta\omega_E(V)}\left[e^{\beta\omega_E(V)} -1\right]^{-2} \:.
\end{align}
For an LJ or ELJ potential we have analytical expressions for both terms through Eqs. (\ref{eq:fccEinstein}), (\ref{eq:PZPVE}) and (\ref{eq:BZPVE}). These equations show that $P_{th}\propto P_{\rm ZPVE},T,\omega_E^{-1}$ and $B_{th}\propto B_{\rm ZPVE},P_{\rm ZPVE}^2,T,\omega_E^{-1}$.  Different formulae for thermal contributions are available from the Debye model which requires the Debye frequency and the Gr\"uneisen parameter.\cite{Zha2004} Using our two formulae we obtain for the ELJ potential at $T=$297 K and $V=2$cm$^3$/mol a thermal pressure component of $P_{th}$= 0.21 GPa and bulk modulus of $B_{th}$= -0.69 GPa ($B_{th1}$= 0.36 GPa, and $B_{th2}$= -1.05 GPa). These are relatively small compared to the measured values of about $P=$110 GPa and $B=$290 GPa at that volume.\cite{Dawaele2020} We find that the $B_{th2}$ term in Eq.(\ref{eq:BfiniteT}) dominates leading to a negative thermal contribution to the bulk modulus, in agreement with the values provided by Zha, Mao and Hemley.\cite{Zha2004} These authors also noted relatively small values for the thermal pressure. The reason for this lies in the small $^4$He mass resulting in a large Einstein frequency $\omega_E$ and small thermal contribution. It explains why the experimental temperature differences for the pressure and bulk modulus between 15 and 297 K are barely visible in Figure \ref{fig:LJnm}(c,d). As shown for neon, the thermal contributions become far more important in the low-pressure regime.\cite{SchwerdtfegerHermann2009} We therefore neglect temperature effects for $^4$He in our discussion because the neglect of higher-body terms contains much larger errors compared to the thermal contributions.

While the qualitative LJ picture shown in Figure \ref{fig:LJnm}(a,b) remains the same for the ELJ potential, the pressure and bulk moduli are a fraction smaller and much closer to the experimental values, that is because the ELJ potential describes the repulsive wall correctly  in contrast to the (12,6) LJ potential. Further improvement requires the inclusion of phonon dispersion and, more importantly, higher $N$-body terms in the interaction potential\cite{Loubeyre1987,Khairallah2008} which become attractive in the short-range.\cite{Ross1986,Ceperley1987,Cohen1996,Ceperley1995,Syshchenko2010,Ceperley2012,Barnes2017,Gang2018} For higher $n$-body forces analytical formulae in terms of lattice sums are unfortunately not available. Moreover, the most accurate three-body potential obtained from ab-initio data by Cencek, Patkowski, and Szalewicz (CPS)\cite{Cencek2007}  is only valid for internuclear distances of $r>$ 3.5 a.u. ($V>$ 2.8 cm$^3$/mol for the fcc structure), and to add to this, the different three-body potentials available\cite{Parish1994,Cohen1996,Wang2006,Cencek2007} lead to quite different results in the short range. Nevertheless, in the valid volume range we calculate a total pressure including three-body effects with the ELJ two-body and CPS three-body potential of 27.3 GPa at $V$=2.954 cm$^3$/mol compared to the experimental value of 35.5 GPa.\cite{Dawaele2020} This underestimation of the pressure at small volumes was also noted by Chang and Boninsegni.\cite{Chang2001} Bulk moduli calculations by Barnes and Hinde show that three-body interactions become very important in the short-range.\cite{Barnes2017} How important the three-body, and higher order, contributions are to the vibrational pressure are topics to be further investigated.

%The difference in Lennard-Jones cohesive energies between the bcc, fcc and hcp phases
\subsection{\label{sec:AppLJ}The difference in Lennard-Jones cohesive energies between the bcc, fcc and hcp phases}

The almost energetically degenerate fcc and hcp phases for the rare gases have been a matter of long standing debate.\cite{Kihara-1952,Wallace1965,Niebel-1974,Waal1991,Ackland2006,Krainyukova-2012} 
We therefore discuss the difference in cohesive energies between the different phases for a $(n,m)$ LJ  potential in more detail. 

\begin{figure}[h!]
\centering
\includegraphics[width=.466\columnwidth]{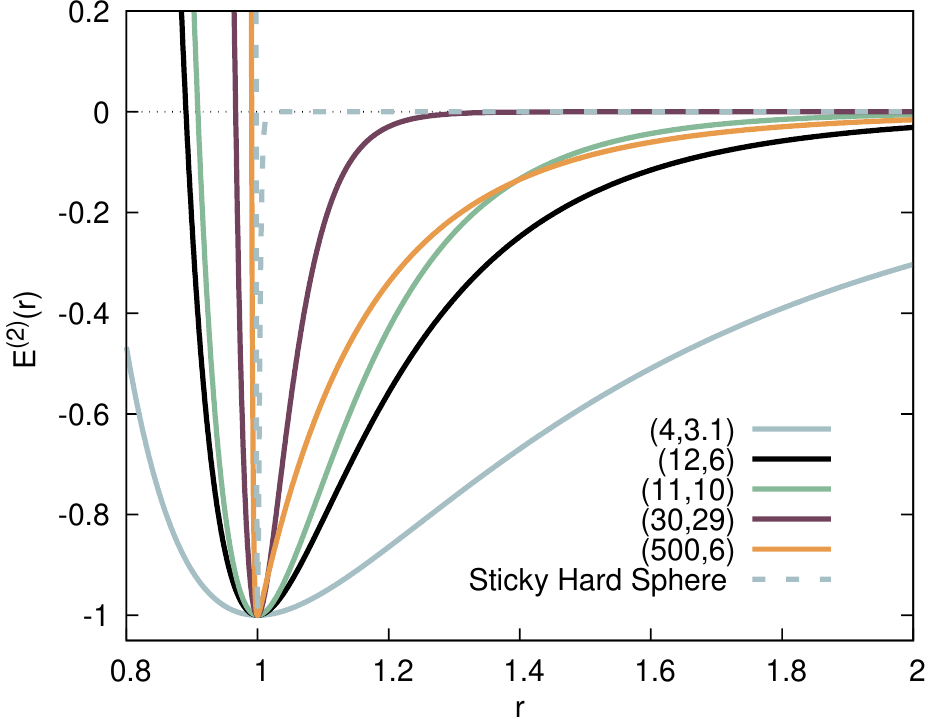}
\includegraphics[width=.45\columnwidth]{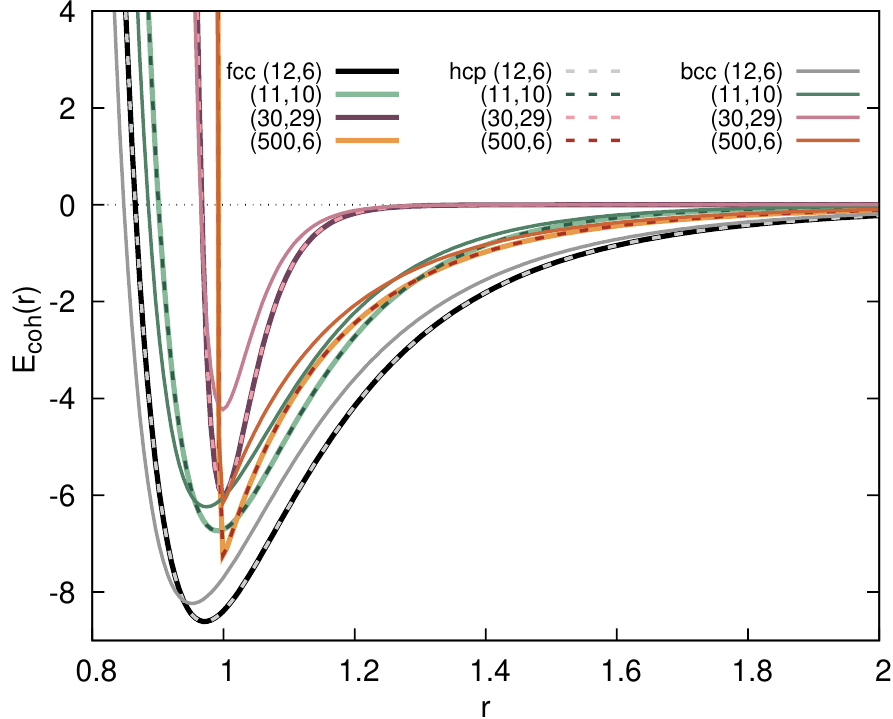}
\includegraphics[width=.45\columnwidth]{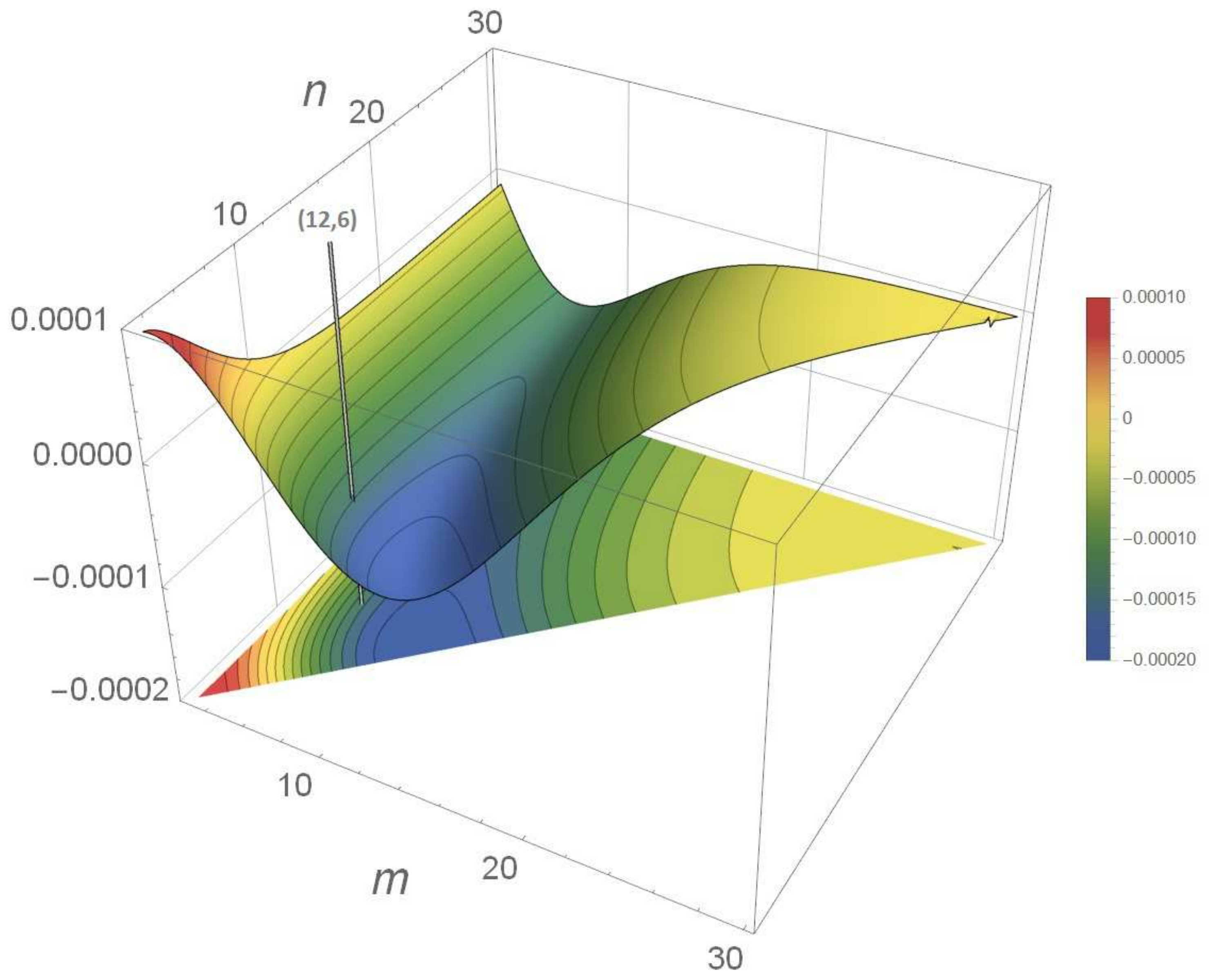}
\includegraphics[width=.45\columnwidth]{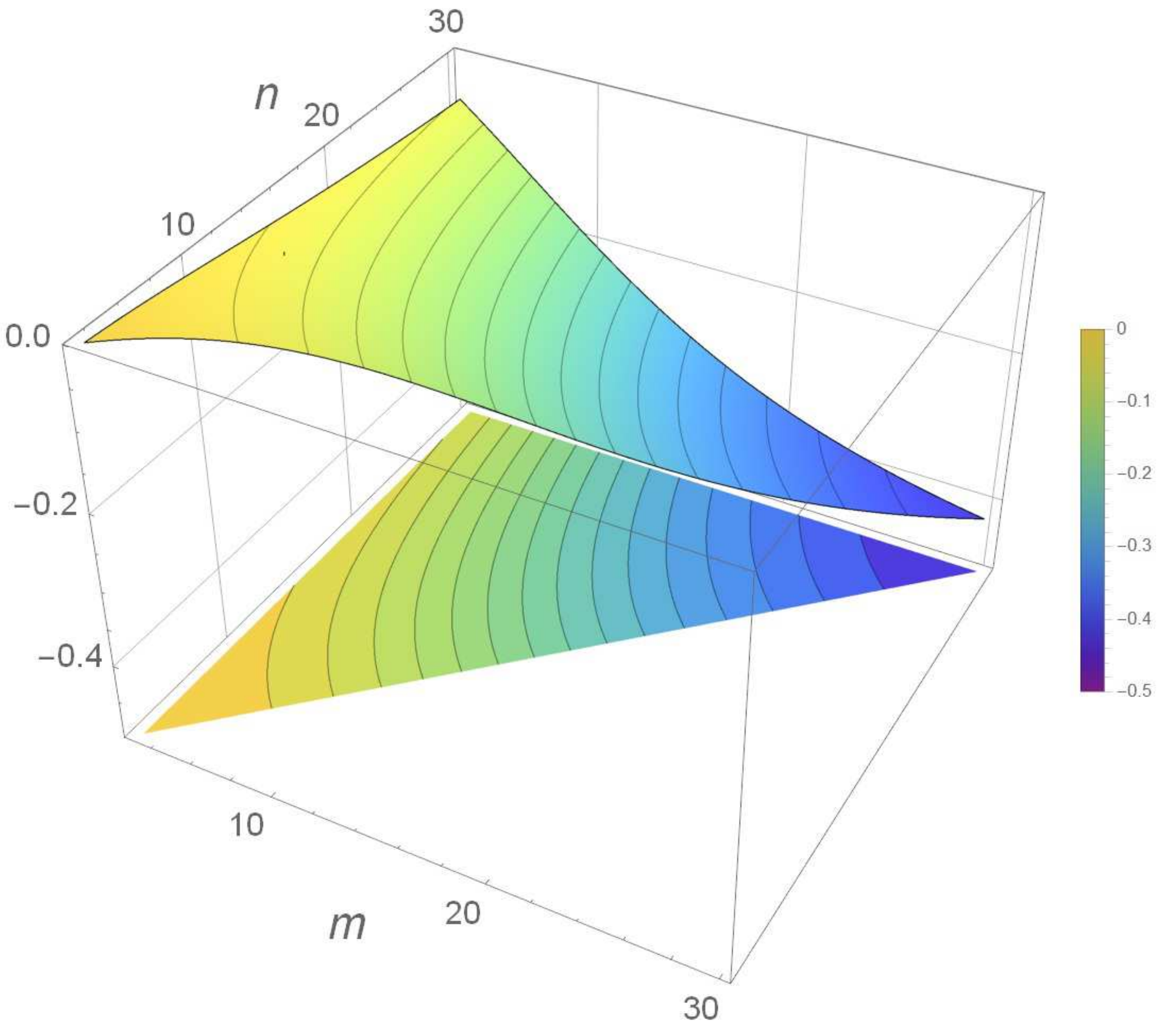}
\caption{(Color online) (a) The interaction potential and (b) the cohesive energy for a range of $n$ and $m$ values of the $(n,m)$ LJ potentials.  Relative difference in cohesive energies (c) $\Delta_{\rm hcp}^{\rm fcc}(n,m)$ between the fcc and hcp phase and (d) $\Delta_{\rm hcp}^{\rm bcc}(n,m)$ between the bcc and hcp phase, for different choices of $(n,m)\in \mathbb{R}_+^2, n>m>3$ of the LJ potential.}
\label{fig:LJnm3D}
\end{figure}

Using Eqs. (\ref{eq:EcLJn-m}) and (\ref{eq:LJdist}) we obtain for the cohesive energy at the minimum nearest neighbor distance,
\begin{equation} \label{eq:EcohLJmn}
E_{\rm LJ}(r_0^{\rm min}) =
\frac{\epsilon}{2(n-m)}\left[mL_n\left(\frac{L_m}{L_n}\right)^{\tfrac{n}{n-m}} - nL_m\left(\frac{L_m}{L_n}\right)^{\tfrac{m}{n-m}}\right] \quad , \quad m<n \:.
\end{equation}
Similar to the minimum neighbor distance which is directly related to the equilibrium distance of the dimer, (see Eq.(\ref{eq:LJdist})), the cohesive energy is only dependent on the binding energy $\epsilon$ of the diatomic and the ratios between LJI coefficients. From this we derive the relative difference in cohesive energies $\Delta_{\rm P1,P2}$ between the two phases P1 and P2,
\begin{align} \label{eq:LJ1}
\Delta&_{\rm P1,P2}(n,m) = 1-\frac{E_{\rm LJ}^{\rm P2}\left(r_0^{\rm P2}\right)}{E_{\rm LJ}^{\rm P1}\left(r_0^{\rm P1}\right)}
=1-\frac{mL_n^{\rm P2}\left(\tfrac{L_m^{\rm P2}}{L_n^{\rm P2}}\right)^{\tfrac{n}{n-m}} - nL_m^{\rm P2}\left(\tfrac{L_m^{\rm P2}}{L_n^{\rm P2}}\right)^{\tfrac{m}{n-m}}}
{mL_n^{\rm P1}\left(\tfrac{L_m^{\rm P1}}{L_n^{\rm P1}}\right)^{\tfrac{n}{n-m}} - nL_m^{\rm P1}\left(\tfrac{L_m^{\rm P1}}{L_n^{\rm P1}}\right)^{\tfrac{m}{n-m}}} \:.
\end{align}
For the (12,6) LJ potential this simplifies to,
\begin{equation} \label{eq:EcohLJ126}
\Delta_{\rm P1,P2}(12,6)=1-\frac{L_{12}^{\rm P1} \left(L_6^{\rm P2}\right)^2} {L_{12}^{\rm P2}\left(L_6^{\rm P1}\right)^2}
\end{equation}
and we obtain $\Delta_{\rm fcc/hcp}(12,6)=-1.00994\times 10^{-4}$ and $\Delta_{\rm bcc/hcp}(12,6)=-4.53763\times 10^{-2}$ using the lattice sums from Ref.\onlinecite{burrows-2020}.
We see that such a potential prefers the hcp structure as correctly analyzed by Kihara and Koba,\cite{Kihara-1952} although fcc is very close in energy.\cite{Loubeyre1987,Loubeyre1988,Schwerdtfeger-2016}  
For a general $(n,m)$ LJ potential allowing for real exponents, one has to introduce unphysical soft potentials of low $(n,m)$ values with $n<5.7$ to stabilize the fcc structure through two-body forces alone as Figure \ref{fig:LJnm3D} shows. The figure also shows that hcp is preferred over bcc through a range of $(n,m)$ values.

The preference for hcp over fcc can easily be explained. Looking at shells of atoms around one arbitrarily-chosen central atom, we find the same numbers of atoms in the first and second shell for the fcc and hcp lattice. Differences only start from the third shell onwards, hcp has two extra atoms at a distance of $\sqrt{8/3}r_0$ that are not present in the fcc structure. Therefore, at a distance of  $\sqrt{8/3}r_0$, the fcc cluster contains 18 atoms while the hcp has already 20 atoms.  The third fcc shell is found at a much larger distance of $\sqrt{3} r_0$ with an additional 24 atoms.\cite{conway-1999,Ackland2006}. This is reflected in the lattice sums, as we obtain the inequality
\begin{equation} \label{LJIdif}
\Delta L_s^{\rm fcc/hcp}=L_s^{\rm fcc}-L_s^{\rm hcp}<0 \:,
\end{equation}
over whole range of real values of $n\in \mathbb{R}_+, n\ge 3$ (also allowing for the singularity at $n=3$).\cite{burrows-2020} In fact, $\Delta L_n^{\rm fcc/hcp}$ has a minimum at $n$=6.2448 with $\Delta L_s^{\rm fcc/hcp}$=-0.00097845 with maximum preference for the hcp structure, which is close to the dispersive $n=6$ term. As the $r^{-6}$ term is the dominant interaction for the first few nearest neighboring shells, this situation does not change if we adopt a more accurate two-body potential.\cite{Schwerdtfeger-2016} This explains that for a simple LJ potential, without inclusion of zero-point vibrational effects, hcp is preferred over the fcc lattice contrary to what is known from experiment.\cite{Schwerdtfeger-2006} The  only exception we find for ultra-soft LJ potentials with small $(n,m)$ values close to the singularity of the lattice sum at $n=3$. Here counting shells further away becomes important.

A special case of the  $(n,m)$ LJ potential is the Sticky Hard Sphere (SHS) potential
\begin{equation}
  V_{SHS}(r)=\begin{cases}
    \infty, &  r<r_e\\
    -\epsilon, &  r=r_e\\    
    0, & r>r_e \end{cases}
\end{equation}
which is reached in the  limit ${n\rightarrow \infty},{m\rightarrow \infty}, n>m$, depicted with the blue dashed line in Figure (\ref{fig:LJnm3D})(a).
The SHS potential does not distinguish between the fcc or hcp phases, i.e. they are energetically degenerate, since both phases have, within this limit, the same packing density, representing the densest possible packings of spheres. In fact, combinations of fcc and hcp layers, so-called Barlow packings also belong to the most dense sphere packings.\cite{Barlow1883} However, such packings have not been observed experimentally, which remains an unresolved problem in the theory of lattice packings.\cite{conway1995all} 
A SHS potential with long-range dispersion can be constructed by using the $(n,6)$ LJ potential with a very large $n$ value, depicted with the orange line in Figure (\ref{fig:LJnm3D})(a) and (b).\cite{Baxter1968,Trombach2018}
In this case the cohesive energy is given by,\cite{Trombach2018}
\begin{equation} \label{SSH}
{\rm lim}_{n\rightarrow \infty} E_{\rm LJ}(r_{0})=-\epsilon \frac{L_m}{2} \left( \frac{r_e}{r_0} \right)^m \:.
\end{equation}
 
%The difference between the fcc and hcp phase for solid argon under pressure
\subsection{\label{sec:Argon}The difference between the fcc and hcp phase for solid argon under pressure}
\begin{figure}
\centering
(a)\includegraphics[width=.45\columnwidth]{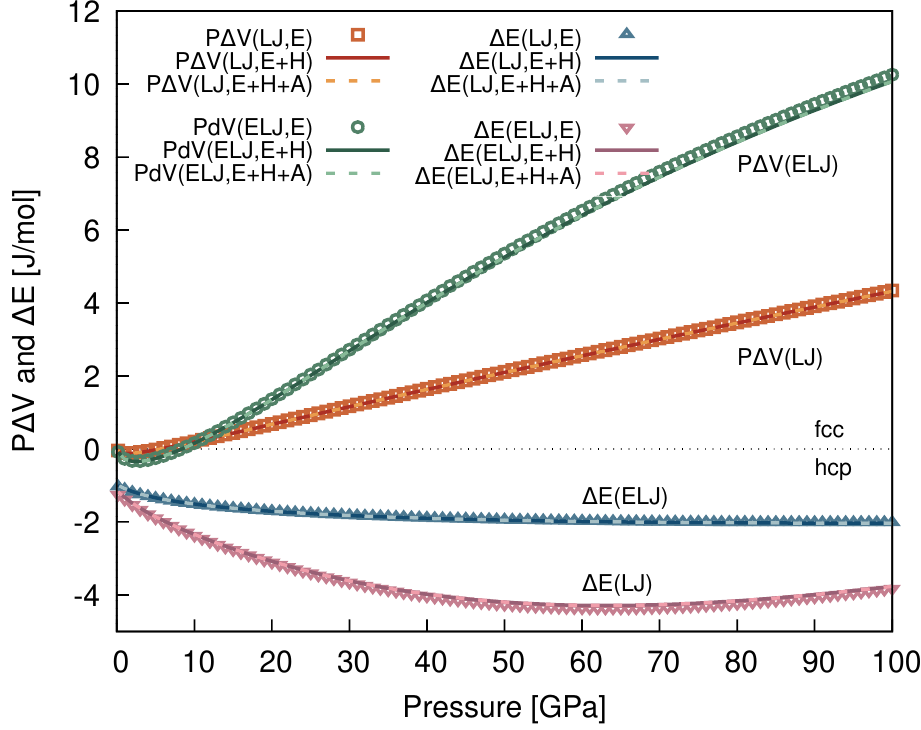}
(b)\includegraphics[width=.45\columnwidth]{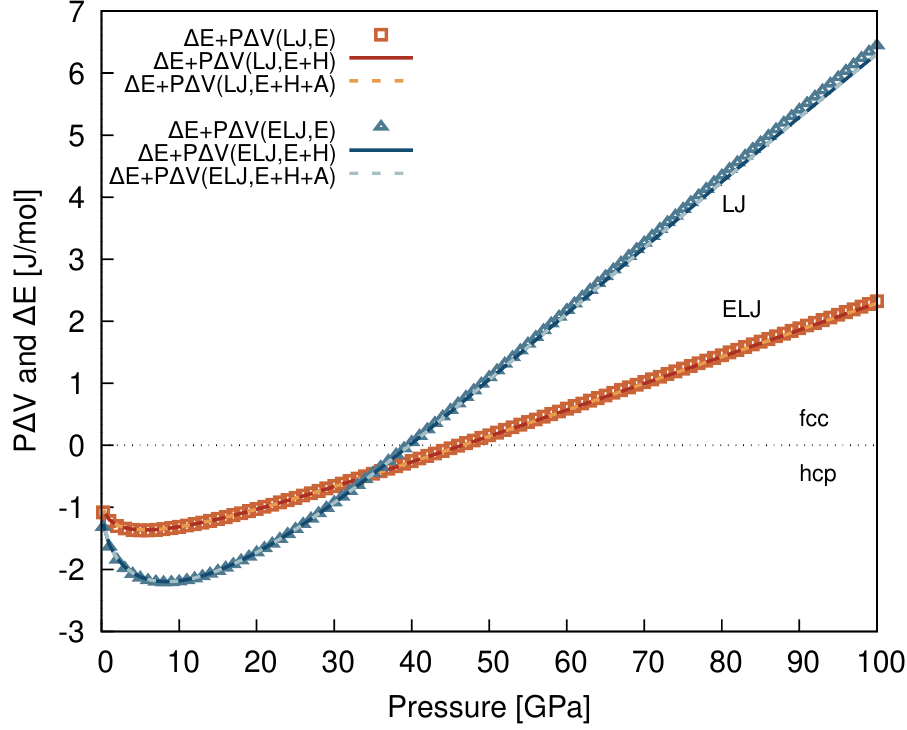}
\caption{Enthalpy difference $\Delta H_{\rm fcc/hcp}(P)$ between fcc and hcp against the pressure $P$ for the LJ and the ELJ potential. Negative values implies that the hcp phase is more stable. (a) lower two curves are  $\Delta E_{\rm hcp/fcc}(P)$ plots, upper two curves $P\Delta V_{\rm hcp/fcc}$ plots. (b) $\Delta E_{\rm hcp/fcc}(P)+\Delta E_{\rm hcp/fcc}(P)$. The individual contributions are cohesive energy expression used only (E), Eqs.(\ref{eq:VolLJ}) and (\ref{eq:PLJ}), harmonic ZPVE added to the cohesive energy expression (H), Eqs.(\ref{eq:VolLJ}), (\ref{eq:PLJ}), (\ref{eq:fccZPE})  and  (\ref{eq:PZPVE}), and finally anharmonicity corrections added (A), Eqs.(\ref{eq:VolLJ}), (\ref{eq:PLJ}), (\ref{eq:fccZPE}), (\ref{eq:fccAZPVE3}),  (\ref{eq:PZPVE}) and (\ref{eq:PAZPVE}).}
\label{fig:twobody}
\end{figure}

In the previous section the difference in cohesive energy between the fcc, hcp and bcc at zero Kelvin was discussed. To compare these phases under pressure, the enthalpy has to be considered instead. The difference in enthalpies between hcp and fcc at constant pressure $P$ at zero Kelvin is,
\begin{equation} \label{LJenthaplpydif}
\begin{aligned}
\Delta H_{\rm hcp/fcc}(P) &= \Delta E_{\rm hcp/fcc}(P) + P\Delta V_{\rm hcp/fcc}(P) \\
&= E_{\rm hcp}[V_{\rm hcp}(P)] - E_{\rm fcc}[V_{\rm fcc}(P)] + P\{V_{\rm hcp}(P)-V_{\rm fcc}(P)\} 
\end{aligned}
\end{equation}
which will be used to determine if the hcp phase persists into the high pressure region for a LJ potential.
Here $E=E^{\rm coh}+E^{\rm ZPVE}+E^{\rm AZPVE}$. For a (12,6) LJ potential relation between pressure and volume is given by Eq.(\ref{eq:PLJ}),
\begin{equation} \label{pressurevolumeLJ}
P(V)=2\epsilon r_e^6\left( r_e^{6}V^{-5} - 2L_6V^{-3} \right)+P^{\rm ZPVE}(V)+P^{\rm AZPVE}(V) \:.
\end{equation}
Even if we neglect vibrational effects, for converting the pressure into volume one has to solve a fifth-order polynomial equation $ax^5+bx^3+c=0$, which according to the Abel-Ruffini theorem has no general analytical solution. If we add vibrational effects both equations become more demanding and we have to get the volume from the pressure through more complicated algebraic equations, which can only be solved by numerical methods. We therefore calculate the volume $V$ from a given pressure $P$ by a two-point interpolation between $(P_1,V_1)$ and $(P_2,V_2)$ using an exponential ansatz,
\begin{equation} \label{PVapprox}
P(V)=Ae^{-aV}  \quad  {\rm with} \quad A=P_1e^{{\rm ln}\left( \frac{P_2}{P_1}\right) \frac{V_1}{V1-V2}}  \quad  {\rm and} \quad a=\frac{{\rm ln}\left( \frac{P_2}{P_2}\right)}{V_1-V_2} \:.
\end{equation}
This results in an iterative process for the volume determination,
\begin{equation} \label{volumefrompressure}
V_1^{(n+1)}=V^{(n)}_1+\frac{\mathrm{ln}\left[ P/P^{(n)}_1\right]}{\mathrm{ln}\left[ P^{(n)}_2/P^{(n)}_1\right]}\left[ V^{(n)}_2-V^{(n)}_1 \right] \:,
\end{equation}
with $V^{(n)}_2=f_nV^{(n)}_1$ with $f_n=1\pm \epsilon$ and $\epsilon \rightarrow 0$ for $n\rightarrow \infty$ ($P^{(n)}_2$ follows from $V^{(n)}_2$). In general, choosing $f_{n+1}=f_n/a$ ($a=5.0$ for example) only five iterations are required to reach computer precision for the volume $V_1^{(n)} \rightarrow V$ at a given pressure $P$. This procedure works well as long as the curve behaves exponential, i.e., in the region where the pressure becomes negative a second-order polynomial fit for $P(V)$ is preferred. We now apply this to the fcc and hcp phase of solid argon at high pressures. The individual contributions for $\Delta H_{\rm fcc/hcp}(P)$ up to pressures of 100 GPa are shown in Figure \ref{fig:twobody}. 

The differences in enthalpies between the fcc and hcp phase are very small (see Figure \ref{fig:twobody}(a)) (in the J/mol range), and this small difference persists up to very high pressures. We also see that at high pressures the $P\Delta V_{\rm hcp/fcc}(P)$ starts to dominate over the $\Delta E_{\rm hcp/fcc}(P) $ term. The almost linear behavior of the $P\Delta V_{\rm hcp/fcc}(P)$ comes from an almost constant value of the volume difference, e.g. $\Delta V_{\rm hcp/fcc}(P)\approx$1.85-1.95$\times 10^{-5}$ cm$^3$/mol in the high pressure range for the LJ potential. Within this model, the hcp phase is preferred at low pressures, while the fcc phase becomes more stable at pressures between 40-50 GPa, Figure \ref{fig:twobody}(b). This is in agreement with Stillinger's analysis\cite{Stillinger2001} who predicts an hcp$\rightarrow$fcc transition for the LJ potential for Ar at a volume ratio of $V/V^{\rm min}$=0.537. This is also the case for the more accurate ELJ potential which we used from Ref.\onlinecite{Pahl_AngewChemIntEd47}. However, this is contrary to experimental findings where a fcc phase is observed at standard conditions,\cite{Barrett-1964,PetersonBatchelderSimmons1966} and a subsequent fcc-to-hcp phase transition occurs at high pressures. In fact, Errandonea et al. observed a broad fcc-to-hcp transition in room temperature X-ray studies extending from 49.6 GPa to an estimated 300 GPa. At the highest pressure of 114 GPa, they determined a ratio of 0.3 for the amount of hcp to fcc.\cite{Errandonea2006} 

We showed recently that the fcc phase is stabilized by phonon dispersion at 0 K.\cite{Schwerdtfeger-2016} As phonon contributions play a lesser role at increased pressures, one can speculate that three- and higher body contributions must be responsible for the phase change to hcp at higher pressures.\cite{troitskaya2019} From a theoretical point of view, to simulate a phase transition with such small enthalpy differences remains a major challenge. If experimental data fitted to many-body potentials, based for example on the embedded atom model, one can obtain more accurate results.\cite{Pechenik2008}

%The Mode Grueneisen Parameter for the Rare Gas Solids
\subsection{\label{sec:Mode-Grueneisen}The Mode Gr{\"u}neisen Parameter for the Rare Gas Solids}

Gr{\"u}neisen stated in 1912 that the parameter $\gamma(V,T)$ is almost independent of volume and temperature and expected to have the same value for elements of similar structure and interaction potential.\cite{Gruneisen1912} An estimate was given by considering nearest neigbor interactions only, (see Ref.\onlinecite{Sherman_1982}) which gives the value of $\gamma=$3.17 for a (12,6) LJ potential,
\begin{equation} 
\gamma = \frac{n+m+1}{6} \:.
\end{equation}
Indeed, the value varies very little for the rare gases from about 2.5 to 2.7,\cite{Tilford1972,fugate1973} but deviates substantially from Gr{\"u}neisen's original estimate. In the following we only consider the volume dependent mode Gr{\"u}neisen parameter, for a discussion on the temperature dependence for the solid and liquid rare gas phases we refer the reader to Refs.\onlinecite{Feldman1967,lurie1973,holste1973,Mausbach2016}.

The Einstein approximation within the LJ model provides a more rigorous insight into the constant value of the mode Gr{\"u}neisen's parameter for the noble gasses. If we substitute Eq.(\ref{eq:LJdist}) into (\ref{eq:GrueneisenLJ1}) we get for the mode Gr{\"u}neisen parameter at distance $r_0=r_0^{\rm min}$,
\begin{equation} \label{eq:Grueneisenrmin}
\begin{aligned}
\gamma_E^{\rm LJ}(r_0^{\rm min})&= \frac{1}{6}\frac{(n+2)(n-1)L_{n+2}L_m-(m+2)(m-1)L_{m+2}L_n}{(n-1)L_{n+2}L_m-(m-1)L_{m+2}L_n}\\
&\mathop{=}\limits_{m=6}^{n=12} \quad \frac{77L_{14}L_6-20L_8L_{12}}{33L_{14}L_6-15L_8L_{12}}
\end{aligned}
\end{equation}
This displays that the mode Gr{\"u}neisen parameter only depends on the type of lattice through their lattice sums. The corresponding values are shown in Table \ref{tab:Grueneisenvalues}. 
\begin{table}[ht!]
\setlength{\tabcolsep}{0.16cm}
\caption{\label{tab:Grueneisenvalues}Dimensionless mode Gr{\"u}neisen parameter $\gamma$ for the four different lattices sc, bcc, fcc and hcp. The LJ values listed are from Eq.(\ref{eq:Grueneisenrmin}) (the LJ value for the simple cubic structure is 2.951916). For the harmonic ({\it h}) and anharmonic part ({\it ah}) we used Eqs.(\ref{eq:GrueneisenEinstein2}) and (\ref{eq:GrueneisenEinstein3}). For He we used the optimized lattice distance $r_0^{\rm min}$ without vibrational effects included as inclusion of ZPVE contributions causes symmetry breaking of the lattice.}
\begin{center}
\begin{tabular}{l|llllll}
Atom & $r_0$(bcc) & $\gamma_E$(bcc) & $r_0$(fcc ) &  $\gamma_E$(fcc) &$r_0$(hcp) &   $\gamma_E$(hcp)\\
\hline
{\it LJ }&  $(\frac{L_{12}}{L_6})^{\frac{1}{6}}r_e$ &  2.991928     & $(\frac{L_{12}}{L_6})^{\frac{1}{6}}r_e$ & 3.014083     & $(\frac{L_{12}}{L_6})^{\frac{1}{6}}r_e$ & 3.014102\\ 
{\it ELJ(h)} \\                                                                                                                       
He 	& 2.84847		&  3.035488	& 2.91126		& 2.767544	& 2.91123		& 2.767651\\
Ne	& 3.09254		&  4.076591	& 3.15380 	& 3.457913 	& 3.15376 	& 3.457846\\
Ar	& 3.66546		&  3.312865	& 3.74303 	& 2.971256 	& 3.74298 	& 2.971366\\
Kr	& 3.87459		&  3.171161	& 3.95843		& 2.857477	& 3.95839		& 2.857453\\
Xe	& 4.20349 	&  3.170670	& 4.29406		& 2.851391	& 4.29401		& 2.851350\\
Rn	& 4.25436		&  2.921074	& 4.35199		& 2.632502 	& 4.35193		& 2.632446\\
Og	& 4.09982		&  2.557463	& 4.20118		& 2.332987	& 4.20112		& 2.333014\\
{\it ELJ(h+ah)} \\
$^4$He 	& 2.84847	& 1.889964	& 2.91126		& 2.030678	& 2.91123		& 2.030698\\
$^{20}$Ne& 3.08478	& 3.068178	& 3.14609		& 2.962300	& 3.14605		& 2.962298\\
$^{40}$Ar	& 3.66501	& 3.163060 	& 3.74258		& 2.886474	& 3.74254		& 2.886454\\
$^{84}$Kr	& 3.87447	& 3.102133	& 3.95831		& 2.817675	& 3.95827		& 2.817654\\
$^{132}$Xe&4.20344& 3.127559	& 4.29400		& 2.826453	& 4.29396		& 2.826419\\
$^{222}$Rn&4.25434& 2.901018	& 4.35198		& 2.621050	& 4.35192		& 2.621010\\
$^{294}$Og&4.09982& 2.549471 	& 4.20118		& 2.328094 	& 4.20112		& 2.328116\\
\hline
\end{tabular}
\end{center}
\end{table}

The LJ $\gamma_E$ value for the fcc lattice is considerably below the value estimated by Gr{\"u}neisen which demonstrates that the summation over the whole lattice is important. Moreover, the $\gamma_E$ values vary only slightly between the different lattices, and the difference between fcc and hcp is miniscule. Table \ref{tab:Grueneisenvalues} also contains ELJ results for the rare gases for both the harmonic and anharmonic approximation at the optimized nearest neighbor distances. These values show that anharmonicity effects play a major role especially for He and Ne.

Table \ref{tab:Grueneisenvalues2} shows the mode Gr{\"u}neisen parameter for the fcc lattice at the experimentally determined nearest neighbor distance in comparison with experimental $\gamma$-values. Considering that phonon dispersion and higher body effects are neglected, our results are in reasonable agreement with experiment. Previous calculations using the Debye model are also in good agreement with experiment.\cite{gupta1969,lurie1973} 
\begin{table}[ht!]
\setlength{\tabcolsep}{0.16cm}
\caption{\label{tab:Grueneisenvalues2}Dimensionless mode Gr{\"u}neisen parameter $\gamma_E$ for the fcc lattice at the experimental nearest neighbor distances\cite{PetersonBatchelderSimmons1966,Batchelder1968,Losee1968,Sears1962} for Ne, Ar, Kr and Xe. Experimental $\gamma$-values are from Refs.\onlinecite{Tilford1972,fugate1973}.}
\begin{center}
\begin{tabular}{l|llllll}
Atom & $r_0^{\rm exp.}$&  $\gamma_E(h)$ &  $\gamma_E(h+ah)$ &  $\gamma_E$(exp.)\\
\hline
$^{20}$Ne		& 3.15681$\pm$0.00006	&3.4757	&2.9866	& 2.51$\pm$0.03\\
$^{40}$Ar		& 3.74779$\pm$0.00006	&2.9869	&2.9011	& 2.7$\pm$0.1\\
$^{84}$Kr		& 3.99223$\pm$0.00007	&2.9592	&2.9126	& 2.67$\pm$0.07\\
$^{132}$Xe	& 4.3358$\pm$0.0004 	&2.9754	&2.9453	& 2.5$\pm$0.1\\
\hline
\end{tabular}
\end{center}
\end{table}

\begin{figure}
\centering
\includegraphics[width=.6\columnwidth]{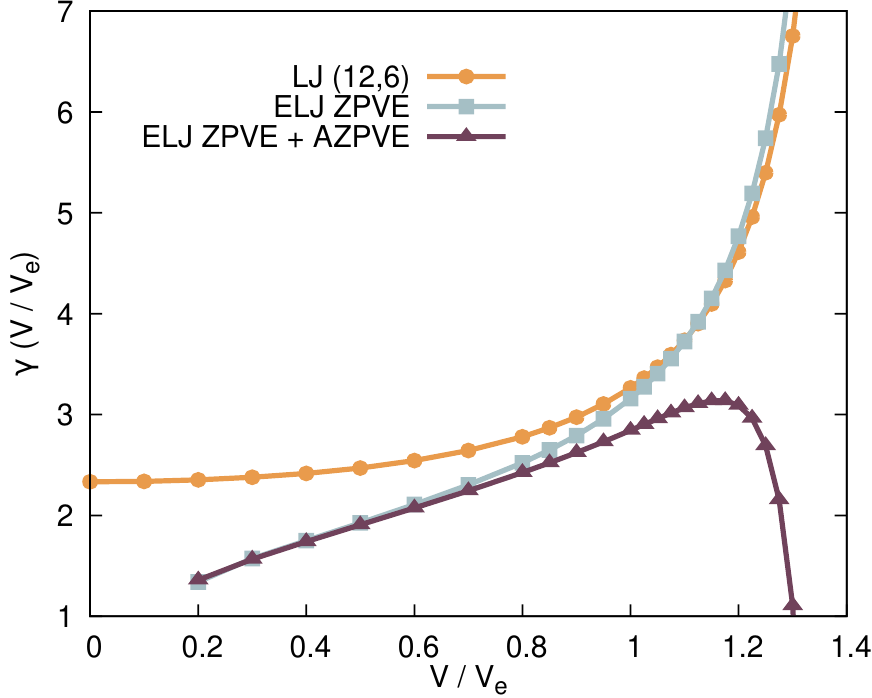}
\caption{Gr{\"u}neisen parameter $\gamma(V/V_e)$ as a function of volume for the LJ potential and for the ELJ of Ne (harmonic and anharmonic). For $r_e$ we used the experimentally derived equilibrium distance of $3.094\pm 0.001$ \AA\cite{WuestMerkt2003} for Ne$_2$ resulting in a volume for solid neon of $V_e=12.612$ cm$^3$/mol.}
\label{fig:Grueneisen}
\end{figure}

Figure \ref{fig:Grueneisen} demonstrates the behavior of $\gamma_E$ for neon over a range of volumes. It shows that, around $V/V_e$ the ELJ and LJ curves are close, but major deviations are observed in the high-pressure regime. Equation (\ref{eq:GrueneisenLJ1}) gives for the high-pressure limit at $\gamma_E(V/V_e=0)=(n+2)/6$ and the point of singularity $\gamma_E(V/V_e)=\infty$ happens at $r_0^{\rm crit}$, Eq.(\ref{eq:Rcrit}), when the denominator in Eq.(\ref{eq:GrueneisenLJ1}) becomes zero. While this behavior has been addressed before,\cite{Karasevskii2003,holzapfel2005} the Einstein approximation provides an analytical explanation.
We observe that in the high-pressure region anharmonicity effects are small, but become important around the equilibrium distance. At distances close to $r_0^{\rm crit}$ the perturbative treatment for anharmonicity effects fails. In this region the mode Gr{\"u}neisen parameter becomes very sensitive to volume changes, which will be especially important for the liquid phase (for a discussion on liquid helium see for example de Souza et al.\cite{de_Souza_2016}).

%Need to check SAMBA: print out all three frequencies, check again cubic force field, does not look right from the output.
%\subsection{\label{sec:appl}Finite temperature results for the Lennard-Jones rare gas solids}
%Corner considered the heat of sublimation for the rare gases.\cite{Corner-1939} ...
%Odile to attend to.

%Summary
\section{\label{sec:summary}Summary}

We derived analytical formulae for the vibrational contributions to the pressure and bulk modulus within the Einstein model, which give us qualitative, yet deep insight into many bulk properties such as the mode Gr{\"u}neisen parameter. The rare gases served as a good starting point to estimate harmonic and anharmonic vibrational contributions to solids. While the LJ potential may be inadequate to model interactions in solids over a large $P(V)$ range, the ELJ potential provides analytical formulae for vibrational effects within the Einstein approximation that are capable for accurately describing two-body interactions over a large volume range. 

There are many open questions in this field. It would be desirable to find approximate analytical formula for the dynamic matrix for an ELJ potential to include phonon dispersion, as well as for the three-body potential such as the Axilrod-Teller-Muto expression\cite{AxilrodTeller1943,Muto1943} or similar expressions which work in the high pressure range. One could, for example, extend the work by Nijboer and deWette\cite{Nijboer1965,Nijboer1971} and use the Terras expansion of quadratic forms in terms of Bessel functions.\cite{terras-1973} Our group is currently trying to resolve these long-standing issues. Specifically, the fcc/hcp phases are very close in energy for the rare gases and the correct treatment of phase diagrams requires phonon dispersion and inclusion of at least three-body forces or even beyond. Especially for helium at high pressures such effects become crucial to correctly predict the $P(V,T)$ and $B(V,T)$ surfaces and phases. Moreover, for helium at low pressures one requires a more accurate quantum treatment.\cite{Ceperley1995}

\begin{acknowledgement}
This work was sponsored by the Marsden Fund administered by the New Zealand Royal Society. We thank Dr. Agnes Dawaele (CEA, DAM, DIF, Arpajon, France) for providing unpublished $P(V)$ data for $^4$He and both Assoc. Prof. Shaun Cooper (Massey University) and Dr. Elke Pahl (Auckland University) for discussions.
\end{acknowledgement}

%\clearpage

\section{\label{sec:ForceField}Appendix: Derivatives of the Extended Lennard-Jones potential in the crystal field}

In order to describe the vibrational motion in a lattice within the Einstein approximation (E), we express the two-body energy of the vibrating atom at position $\vec{r}$ in the ELJ field of all other (fixed) atoms $i\in\mathbb{N}$ positioned at $\vec{r}_i = (x_i, y_i, z_i)^\top $ as
\begin{equation} \label{eq:LELJ}
E\left( \vec{r} \right) = \sum_{i=1}^{\infty}\sum_{n>3} c_n|\vec{r}-\vec{r}_i|^{-n} \:.
\end{equation}
with, $r_i=[(x-x_i)^2+(y-y_i)^2+(z-z_i)^2]^{\frac{1}{2}}$  the distance between the central vibrating atom and the other atoms $i$.   
This extends the work of Corner\cite{Corner-1939} and Wallace\cite{Wallace1963,Wallace1964,Wallace1965} to the terms in the ELJ potential.

A Taylor expansion around the origin is defined by,
\begin{equation} \label{eq:Taylor}
E\left( \vec{r} \right) = \sum_{k=0}^\infty \frac{1}{m!} \left( \vec{r}\cdot \vec{\nabla} \right)^m E\left( \vec{r} \right)\vert_{\vec{0}} \:.
\end{equation}
All derivatives in cartesian coordinates up to 4th order with respect to the atom moving around the origin may now be derived
\begin{equation} \label{eq:1PxLELJ}
F_x = \left.\frac{\partial E} {\partial x}\right\vert_{\vec{0}} =
\sum_{i,n} n c_n x_i r_i^{-n-2}
\end{equation}
\begin{equation} \label{eq:2PxyLELJ}
F_{xy} = \left.\frac{\partial^2 E}{\partial x \partial y}\right\vert_{\vec{0}} =
\sum_{i,n} n\left( n+2\right) c_n x_i y_i r_i^{-n-4} \:,
\end{equation}
\begin{equation} \label{eq:2PxxLELJ}
F_{xx} = \left.\frac{\partial^2 E}{\partial x^2}\right\vert_{\vec{0}} =
\sum_{i,n} n c_n r_i^{-n-4} \left[\left( n+2\right) x_i^2 - r_i^2 \right] \:,
\end{equation}
\begin{equation} \label{eq:3PxxxLELJ}
F_{xxx} = \left.\frac{\partial^3 E}{\partial x^3}\right\vert_{\vec{0}} = 
\sum_{i,n} n \left( n+2 \right) c_n x_i r_i^{-n-6} \left[\left( n+4\right) x_i^2 - 3r_i^2 \right] \:,
\end{equation}
\begin{equation} \label{eq:3PxxyLELJ}
F_{xxy} = \left.\frac{\partial^3 E}{\partial x^2\partial y}\right\vert_{\vec{0}} = 
\sum_{i,n} n \left( n+2 \right) c_n y_i r_i^{-n-6} \left[\left( n+4\right) x_i^2 - r_i^2 \right] \:,
\end{equation}
\begin{equation} \label{eq:3PxyzLELJ}
F_{xyz} = \left.\frac{\partial^3 E}{\partial x\partial y\partial z}\right\vert_{\vec{0}} = 
\sum_{i,n} n\left( n+2\right) \left( n+4\right) c_n x_i y_i z_i r_i^{-n-6} \:,
\end{equation}
\begin{equation} \label{eq:4PxxxxLELJ}
F_{xxxx} = \left.\frac{\partial^4 E}{\partial x^4}\right\vert_{\vec{0}} = 
\sum_{i,n} n \left( n+2 \right) c_n r_i^{-n-8} \left[ 3r_i^{4} - 6\left( n+4\right) x_i^2 r_i^2 + \left( n+4 \right) \left( n+6 \right) x_i^4 \right] \:,
\end{equation}
\begin{equation} \label{eq:4PxxxyLELJ}
F_{xxxy} = \left.\frac{\partial^4 E}{\partial x^3\partial y}\right\vert_{\vec{0}} = 
\sum_{i,n} n \left( n+2 \right) \left( n+4 \right) c_n x_i y_i r_i^{-n-8} \left[\left( n+6\right) x_i^2 - 3 r_i^2 \right] \:,
\end{equation}
\begin{equation} \label{eq:4PxxyyLELJ}
F_{xxyy} = \left.\frac{\partial^4 E}{\partial x^2\partial y^2}\right\vert_{\vec{0}} =
\sum_{i,n} n \left( n+2 \right) c_n r_i^{-n-8} \left[ r_i^{4} - \left( n+4\right) \left( x_i^2 + y_i^2\right) r_i^2 + \left( n+4 \right) \left( n+6 \right) x_i^2 y_i^2 \right] \:,
\end{equation}
\begin{equation} \label{eq:4PxxyzLELJ}
F_{xxyz} = \left.\frac{\partial^4 E}{\partial x^2\partial y\partial z}\right\vert_{\vec{0}} =
\sum_{i,n} n \left( n+2 \right) \left( n+4 \right) c_n y_i z_i r_i^{-n-8} \left[\left( n+6\right) x_i^2 - r_i^2 \right] \:.
\end{equation}
From Eq.(\ref{eq:2PxxLELJ}), we derive the Laplacian $\Delta E$ with respect to our vibrating atom,
\begin{equation}
\Delta E\vert_{\vec{0}} =  \left( F_{xx} +F_{yy} +F_{zz}\right)\vert_{\vec{0}} = \Tr\{F\}\vert_{\vec{0}} = \sum_{i,n} n\left( n-1\right)c_n r_i^{-n-2}
\end{equation}
The cubic lattices sc, bcc and fcc belong to the local $O_h$ point group. If we rotate the orthogonal coordinate system such that $F$ is diagonal\footnote{This normal coordinate system is identical to the orthogonal coordinate system commonly used for the cubic Bravais lattices.} we have $F_{xx}=F_{yy}=F_{zz}$ because of $O_h$ symmetry (not for hcp as already mentioned).\cite{Wallace1965} Thus we obtain
\begin{equation} \label{eq:2PxxFCCELJ}
F_{xx}^{\rm c}\vert_{\vec{0}} = \frac{1}{3}\Delta E\vert_{\vec{0}} =\frac{1}{3}\sum_{i,n} n\left( n-1\right)c_n r_i^{-n-2} \:,
\end{equation}
where $(c)$ stands for one of the cubic lattices. In this case we obtain also simple relationships for the quartic force constants; $F_{xxyy}^{\rm c}=F_{yyzz}^{\rm c}=F_{xxzz}^{\rm c}$ and $F_{xxxx}^{\rm c}=F_{yyyy}^{\rm c}=F_{zzzz}^{\rm c}$.\cite{Wallace1965} Further we have $F_{xxxy}^{\rm c}=F_{xxyz}^{\rm c}=0$. Because $O_h$ contains inversion symmetry we also have $F_{x^iy^jz^k}^{\rm c}=0$ for any odd combination $(i+j+k$), for example $F_{x}^{\rm c}=0$, $F_{xxx}^{\rm c}=0$, $F_{xyy}^{\rm c}=0$, and $F_{xyz}^{\rm c}=0$. Thus, all odd derivatives vanish and for these lattices we only have to consider the quartic force constants for the anharmonicity correction (see below). Using these symmetry relations we can further simplify the two important (non-zero) quartic force constants for the cubic lattices,
\begin{equation} \label{eq:4PxxxySLELJ1}
F_{xxxx}^{\rm c}\vert_{\vec{0}} = \frac{1}{3} \left( F_{xxxx} + F_{yyyy} + F_{xxxx} \right)\vert_{\vec{0}}
\end{equation}
and
\begin{equation} \label{eq:4PxxyySLELJ1}
F_{xxyy}^{\rm c}\vert_{\vec{0}} = \frac{1}{3} \left( F_{xxyy} + F_{xxzz} + F_{yyzz} \right)\vert_{\vec{0}} \:.
\end{equation}
which gives
\begin{equation} \label{eq:4PxxxySLELJ2}
F_{xxxx}^{\rm c}\vert_{\vec{0}} =
\frac{1}{3} \sum_{i,n} n \left( n+2 \right) c_n r_i^{-n-8} \left[ \left( n+4 \right) \left( n+6 \right) \left( x_i^4 + y_i^4 + z_i^4 \right) - 3\left( 2n+5\right) r_i^{4} \right]
\end{equation}
and
\begin{equation} \label{eq:4PxxyySLELJ2}
F_{xxyy}^{\rm c}\vert_{\vec{0}} =
\frac{1}{3} \sum_{i,n} n \left( n+2 \right) c_n r_i^{-n-8} \left[ \left( n+4 \right) \left( n+6 \right) \left( x_i^2 y_i^2 + y_i^2 z_i^2 + x_i^2 z_i^2 \right) - \left( 2n+5\right) r_i^{4} \right] \:.
\end{equation}
No further simplification is possible. However, we can combine Eqs.(\ref{eq:4PxxxySLELJ2}) and (\ref{eq:4PxxyySLELJ2}) and we obtain
\begin{equation} \label{eq:4PxxxySLELJ3}
F_{xxxx}^{\rm c}\vert_{\vec{0}} + 2F_{xxyy}^{\rm c}\vert_{\vec{0}} =
\frac{1}{3} \sum_{i,n} \left( n+2 \right) \left( n+1 \right) n \left( n-1 \right) c_n r_i^{-n-4} \:.
\end{equation}
%which will be used in the next section together. As a more general result we obtain from the fact that the force $\vec{F}=\vec{0}$ that {\it at constant pressure within the Einstein approximation the fcc lattice does not distort above a critical volume $V>V_{\rm crit}$ where $tr(F)>0$} (see below), even at very high pressures, and therefore represents a true (local) minimum at constant pressure within a 2-body approximation. In the following we neglect the notation about the origin used in eqs. (\ref{eq:1PxLELJ}) to (\ref{eq:4PxxyzLELJ}) and understand that for example $F_x \equiv F_x\vert_{\vec{0}}$.

%\begin{suppinfo}
%\end{suppinfo}

\bibliography{references}

%% graphical table of contents.
%%%%%%%%%%%%%%%%%%%%%%%%%%%%%%%%%%%%%%%%%%%%%%%%%%%%%%%%%%%%%%%%%%%%%

%\begin{tocentry}
%
%Some journals require a graphical entry for the Table of Contents.
%This should be laid out ``print ready'' so that the sizing of the
%Inside the \texttt{tocentry} environment, the font used is Helvetica
%8\,pt, as required by \emph{Journal of the American Chemical
%The surrounding frame is 9\,cm by 3.5\,cm, which is the maximum
%permitted for  \emph{Journal of the American Chemical Society}
%graphical table of content entries. The box will not resize if the
%content is too big: instead it will overflow the edge of the box.
%
%This box and the associated title will always be printed on a
%separate page at the end of the document.
%
%\end{tocentry}

\end{document}